%% file: CSI_Nexmon_localization.tex
\documentclass[journal,comsoc]{IEEEtran}
\IEEEoverridecommandlockouts

\usepackage{cite}
\usepackage{amsmath,amssymb,amsfonts}
\usepackage{mathrsfs}
\usepackage{algorithm,algpseudocode}
\usepackage{graphicx}
\usepackage{textcomp}
\usepackage{xcolor}
\usepackage{subfig}
\usepackage{multirow}
\usepackage{comment}
\graphicspath{{./figures/}}

\def\BibTeX{{\rm B\kern-.05em{\sc i\kern-.025em b}\kern-.08em
    T\kern-.1667em\lower.7ex\hbox{E}\kern-.125emX}}
\begin{document}

\title{CRISLoc: Reconstructable CSI Fingerprinting for Indoor Smartphone Localization
\thanks{Zhihui Gao and Yunfan Gao contribute equally to this work.}}

\author{
	Zhihui~Gao,~Yunfan~Gao, ~Sulei~Wang, ~Dan~Li,~Yuedong~Xu
\thanks{
	\noindent Zhihui Gao, Yunfan Gao, Sulei Wang, Dan Li and Yuedong Xu are with School of Information Science and Technology, Fudan University, Shanghai 200433, China.
	Email: \{zhgao16, gaoyf15, wangsl16, lidan,  ydxu\}@fudan.edu.cn
}

}
\maketitle

\begin{abstract}

Channel state information (CSI) based fingerprinting for WIFI indoor localization has attracted lots of attention very recently. The frequency diverse and temporally stable CSI  better represents the location dependent channel characteristics than the coarse received signal strength (RSS). However, the acquisition of CSI requires the cooperation of access points (APs) and involves only data frames, which imposes restrictions on real-world deployment. In this paper, we present CRISLoc, the first CSI fingerprinting based localization prototype system using ubiquitous smartphones. CRISLoc operates in a completely passive mode, overhearing the packets on-the-fly for his own CSI acquisition. The smartphone CSI is sanitized via calibrating the distortion enforced by WiFi amplifier circuits. CRISLoc tackles the challenge of altered APs with a joint clustering and outlier detection method to find them. A novel transfer learning approach is proposed to reconstruct the high-dimensional CSI fingerprint database on the basis of the outdated fingerprints and a few fresh measurements, and an enhanced KNN approach is proposed to pinpoint the location of a smartphone. Our study reveals important properties about the stability and sensitivity of smartphone CSI that has not been reported previously. Experimental results show that CRISLoc can achieve a mean error of around 0.29m in a $6m\times 8m$ research laboratory. The mean error increases by 5.4 cm and 8.6 cm upon the movement of one and two APs, which validates the robustness of CRISLoc against environment changes. 

\end{abstract}


\input{1_introduction}

\input{7_related.tex}
\input{2_preliminaries.tex}
\input{3_overview}

\input{4_1_preprocessing.tex}
\input{4_2_detection.tex}
\input{4_3_transfer.tex}
\input{4_4_matching.tex}

\input{5_setup.tex}

\input{6_1_matching.tex}
\input{6_2_detection.tex}
\input{6_3_transfer.tex}
\input{6_4_overall.tex}

\input{8_conclusion.tex}

\input{appendix}

\bibliographystyle{IEEEtran}
\bibliography{reference}

\end{document}

%% file: 1_introduction.tex
\section{Introduction}

\IEEEPARstart{W}{iFi} indoor localization has witnessed tremendous progress in the past decade owing to 
the pervasive deployment of wireless local area networks (WLANs). 
For instance, Bahl Padmanabhan adopted the Euclidean distance as the matching rule to compare the received RSS vector 
with the stored fingerprints \cite{832252}. Authors in \cite{youssef2005horus} and \cite{tian2017improve} took account of temporal-spatial patterns when 
constructing the fingerprint database.

A recent trend of WiFi fingerprinting is to replace RSS by channel state information (CSI) that represents the channel properties over all the subcarriers at the frequency domain of Orthogonal Frequency Division Multiplexing (OFDM) systems \cite{xiao2012fifs}\cite{wang2015deepfi}\cite{wang2015phasefi}. 
In each packet transmission, a vector of complex CSI values are obtained, instead of a single RSS value. 
The CSI amplitude is not only more temporally stable, but also more representative in terms of the feature of a location 
owing to frequency diversity on different subcarriers. 
However, CSI fingerprinting based localization is obstructed by two practical issues. 
One is the \textit{practical CSI acquisition}. Existing studies rely on Intel 5300 CSI Tool \cite{Halperin_csitool} and Atheros CSI Tool \cite{Xie:2015:PPD:2789168.2790124} toolkits to 
extract CSI from received data packets. 
The operation of CSI Tool either requires the successful connection to each AP or the hardcoded MAC address for passive 
monitoring so that the conducting the site survey is time consuming with many APs. It does not function when the surrounding APs are password protected or operated by a 
WLAN controller for roaming management. 
The other is the \textit{variability of AP deployment}. When a subset of APs malfunction or have been replaced, these changes 
should be detected automatically, and the fingerprint database should be updated accordingly. 
Hence, to make the best of CSI, the indoor localization system should figure out a convenient way of acquiring CSI and 
reconfigure the fingerprint database against the change of AP deployment.

In this paper, we present CRISLoc, the first WiFi CSI fingerprinting localization system using ubiquitous \textit{smartphones}. 
The practical advantages of CRISLoc are twofold. Firstly, CRISLoc can operate in a completely passive mode, overhearing the 
packets on-the-fly for CSI acquisition when the smartphone is not allowed to access APs or different APs sharing the same SSID. 
In addition, CRISLoc can acquire CSI from data frames, acknowledgement (ACK) frames\footnote{The ACK frame is a frame that acknowledges receiving a data frame in MAC layer transmission.} and beacons through smartphones, 
thus providing more freedoms of implementation than previous systems. 
In a nutshell, CRISLoc pushes the conceptual CSI fingerprinting 
closer toward real-world deployment. Secondly, CRISLoc is able to detect the variation of CSI fingerprints when 
one or more APs change their positions. The CSI fingerprints of the altered APs 
are reconstructed using advanced machine learning techniques. Hence, CRISLoc has the potential to achieve both high 
localization accuracy and resilience against environmental changes.

Designing CRISLoc is technically challenging, mixed with profound observations of CSI in smartphones. 
Our first obstacle is the CSI calibration. CRISLoc constructs the fingerprint database 
using a newly developed smartphone CSI toolkit named Nexmon \cite{schulz2018shadow}. However,  the raw CSI amplitudes 
cannot be directly used because the automatic gain control (AGC) function distorts the measured amplitude of 
the received signal. We calibrate the measured CSI amplitudes by removing AGC, and design a couple of filters to 
dispose of unstable subcarriers and abnormal frames. After the pre-processing, CRISLoc 
can use around fifty subcarriers, nearly two times that of Intel 5300 CSI Tool, and the CSI amplitude measurement 
is more stable over time. 

Our second obstacle is the detection of the altered APs as well as the reconstruction of their CSI fingerprints. 
When the localization is carried out using different partitions of the AP set, the estimated positions are likely to huddle 
together if none of the APs in these subsets are redeployed, and are prone to being scattered otherwise. 
The estimated locations may fall in several 
clusters of comparable sizes due to estimation errors, making the direct clustering analysis very fishy. We develop a joint clustering and outlier approach 
to gauge the sizes of largest two clusters that detects the altered APs with both high precision and high recall. 
The CSI fingerprints of altered APs, though becoming obsolete, reflect the room layout, the path loss pattern and the 
spatial correlation. Therefore, we propose to exploit \textit{transfer learning} to distill the knowledge gained from 
the outdated fingerprints other than discarding them. To this goal, a novel optimization framework is formulated 
with the target of finding a transform matrix that projects both the outdated and the fresh CSI data into a subspace 
where the distributions of high-dimensional CSI data match well.

Our third obstacle is to cope with corner cases in indoor localization. An interesting observation regarding the spatial pattern 
of CSI amplitudes is that they are relatively less sensitive to the increase of propagation distance compared with RSS. 
Therefore, choosing as many neighbors in KNN as possible does not yield a more accurate CSI localization, especially when the 
target smartphone is placed at the corners of a room. We proposed an edge enhanced $k$-nearest neighbors (EEKNN) 
method that automatically adjusts the number of neighbors  and their corresponding location-dependent weights. 

In summary, our main contributions are as follows.
\begin{itemize}

\item To the best of our knowledge, CRISLoc is the first CSI fingerprinting localization system using off-the-shelf smartphones. 

\item We design a suit of methods to sanitize CSI data, encompassing the cancellation of automatic gain control 
and the filtering of unstable subcarriers and frames. 

\item We design a joint clustering and outlier detection approach to find the altered APs, and develop a novel transfer learning 
approach to reconstruct their CSI fingerprints.

\item We point out the imperfection of CSI as the location feature, and present an enhanced KNN approach 
to improve the localization accuracy of corner cases. 

\end{itemize}

The remainder of this paper is organized as follows. Several works related to indoor localization and discussions are displayed in Section \ref{sec: related}. Section \ref{sec:preliminary} introduces several preliminaries. Section \ref{sec:overview} states the overall advantages of CRISLoc and its system diagram. Section \ref{sec: systemdesign} and Section \ref{sec:transfer} present the detailed system design and the CSI fingerprint reconstruction. In Section \ref{sec: setup}, we introduce the setup of our experimental environment and claim some metrics and Section \ref{sec: results} shows an exhaustive evaluation of CRISLoc. Finally, Section \ref{sec: conclusion} summarizes the properties of CRISLoc.

%% file: 7_related.tex
\section{Related Work}
\label{sec: related}
\subsection{Measurements}
As the accessible patterns on smartphones, RSS and {\color{blue}signal-to-noise ratio} are used in traditional Wi-Fi fingerprint \cite{832252}\cite{youssef2005horus}\cite{he2015wi}\cite{hossain2011ssd}\cite{laoudias2013crowdsourced}\cite{tian2017improve}. Recently, He and Gary's work concentrated on how WLAN chips on smartphones influence the RSS across different devices\cite{he2015wi} and methods have been purposed to calibrate the heterogeneity of smartphones \cite{hossain2011ssd}\cite{laoudias2013crowdsourced}.
With the introduction of CSI Tool \cite{Halperin_csitool}, Wi-Fi fingerprint localization such as \cite{xiao2012fifs} achieved higher accuracy because of CSI's high-dimensional properties. CSI amplitudes and phases are separately utilized for localization with deep learning in \cite{wang2015deepfi} and \cite{wang2015phasefi}. However, CSI Tool can only be implemented on computers, thus making CSI-based localization less feasible in daily life than in experimental settings.\\
\subsection{Matching Rule}
Recent research has focused on the efficiency of various of matching rules \cite{muller2014field}. Furthermore, different from RSS, CSI is a vector as fingerprints, which requires more sophisticated matching rules. A number of positioning algorithms that match the online measurement with offline fingerprints have been employed. $k$-nearest neighbors (KNN), support vector machine (SVM), and maximum a posteriori estimation (MAP) were used in \cite{Sen:2012:YFM:2307636.2307654}, \cite{1297088}, and \cite{xiao2012fifs} respectively. Besides, the authors in \cite{7438932}\cite{wang2015deepfi}\cite{wang2015phasefi} made use of deep learning approach to perform localization. A matching rule that only utilizes the nearby AP beacons and thus decreases the matching computation complexity is proposed in \cite{8667640}.
\\
\subsection{Site-survey Overhead Reduction}
One weakness of fingerprinting based localization is that fingerprint collection is time-consuming and expensive. In order to lower the site survey overhead, various researches have been conducted. The authors in \cite{Ferris:2007:WUG:1625275.1625675} determined the latent-space locations of unlabeled RSS data by the Gaussian Process Latent Variable Model (GPLVM). Walkie-Markie {\cite{180297}} exploited WiFi landmarks and crowd-sourced trajectory information to automatically build internal pathway maps of buildings. 
In \cite{nabati2020using}, virtual synthetic fingerprints was created by generative adversarial networks (GAN). Such virtual fingerprints are utilized together with other fingerprints collected in reality. The authors in \cite{8736717} introduced a classic GraphSLAM framework to match the collected fingerprints and the the ground truth position, facilitating the process of building database and matching fingerprints.
\\
\indent
Given that signal environment changes over time and regular site survey is required to maintain localization accuracy, many researchers aim to relieve its high cost by auto-reconstructing the fingerprints. The authors in \cite{zheng2008transferring} employed a linear regression model to encode the correlations between RSS at RPs and non-RPs based on the initial fingerprint, and then updated non-RPs' signal strength with the newly collected data at RPs. LAAFU \cite{Cluster} updated the fingerprint using Gaussian process regression and path loss model. Yet, this approach can be applied to CSI only if we use the power of CSI as fingerprints instead of CSI vectors. This is because CSI amplitude of each subcarrier itself does not follow the path loss model. Information conveyed by different subcarriers is then lost and the accuracy of localization significantly decreases.\\
\indent
Transfer learning has been used in site-survey reduction as well. The work in \cite{pan2008transferring} learned a low-dimensional manifold shared between RSS collected in different areas, which enabled the localization model to be transferred from one area of a building to another. The authors in \cite{8030076} learned the distance metrics that gathers RSS vectors in the same spatial cluster and separates RSS vectors in different spatial clusters based on a well-built fingerprint. The learned metrics reduced the required number of site-survey points for constructing an accurate fingerprint database. Transfer kernel learning (TKL) was used in \cite{7925444} to match the outdated and updated RSS distribution in the reproducing kernel Hilbert space. The kernel was then taken as the input of the localization model, which provided accurate location estimation despite environmental dynamics. In \cite{6751384} and \cite{pan2010domain}, MMD also serves as the metrics in transfer learning, based on which fingerprints are reconstructed in \cite{6751384}.\\

\subsection{Other Localization Approaches and Discussion}
Angle-of-arrival (AoA)-based and time-of-flight (ToF)-based solutions are the two other localization methods. These approaches were developed based on the MUSIC algorithm \cite{schmidt1986multiple} and the first implementation is ArrayTrack \cite{xiong2013arraytrack}. By using frequency-agile wireless networks, ToneTrack \cite{xiong2015tonetrack} increased the effective bandwidth and improved localization accuracy. However, due to the blocked direct path and the hardware imperfection, the accuracy of these solutions was usually low in real-world applications.\\
\indent
Apart from WiFi signals, some other wireless communication services were utilized for localization as well. Radio frequency identification (RFID)-based localization such as \cite{Jiang:2018:ORT:3207947.3208004} achieved high accuracy, yet requiring additional RFID tags. Bluetooth Low Energy (BLE) beacon based localization \cite{orujov2018smartphone} is portable in computation complexity but it relies heavily on the path-loss model to apply to a complex environment. Accelerator and gyroscope based localization \cite{Wang:2012:NNW:2307636.2307655}\cite{ashraf2019application} uses dead reckoning method, which suffers great accumulative errors.\\
\indent
Some assumptions are made in CRISLoc's implementation. First, all the CSI is collected in the same direction. The CSI subtly differs when the smartphone heads differently. To solve this problem, a more exhaustive database including fingerprints in each direction is required and the inertial gyroscope is exploited to distinguish the smartphone's actual direction.
Second, the smartphones stay still when collecting CSI. CSI might suffers Doppler effect when the smartphones move too fast. In addition, users' moving fast leads to insufficient time to collect enough CSI samples, which influences CRISLoc's performance, such as the invalid Mahalanobis distance in frame filtering.

%% file: 2_preliminaries.tex
\section{Preliminaries}
\label{sec:preliminary}

In this section, we introduce the basic physical concepts of WiFi indoor localization, traditional methods of fingerprint matching and the machine learning approaches 
that are pertinent to this work. 

\subsection{Received Signal Strength and Channel State Information}

Received signal strength (RSS) indicates the power of a signal received at the physical layer in the unit of decibel. 
In general, the farther the receiver is away from the transmitter, the lower RSS will be.
A widely adopted path loss model, namely wall attenuation factor (WAF) model \cite{seidel1992914}, that captures the 
signal attenuation at GHz frequency band in indoor environments takes the following form:
\begin{equation}
\small
    \label{eq: 1}
    P(d)=P(d_0)-10n\log(\frac{d}{d_0})-\left\{
    \begin{array}{ll}
        n_W\times WAF & {n_W < C}\\
        C\times WAF & {n_W\geq C}
    \end{array} \right.
\end{equation}
where $n$ is the measured attenuation rate, $C$ is a predefined threshold, and $n_W$ is the number of walls between the transmitter and the receiver. $P(d_0)$ is the RSS at a reference position $d_0$ meters away from the transmitter, and $P(d)$ is 
that of the receiver $d$ meters apart from the transmitter. Both $P(d)$ and $P(d_0)$ are in the unit of dBm.
The path loss models provide a rigorous approach to quantify the impact of transmission distance and environmental parameters 
on the signal strength capacity 
of wireless links. Nevertheless, existing models cannot precisely capture the complex signal attenuation, or tell the subtle differences between RSS values of adjacent locations.

Channel state information (CSI) describes the channel properties of a communication link, especially how 
a signal propagates from the transmitter to the receiver and represents the combined effect of 
scattering, fading, and power decay with distance \cite{tulino2005impact}. In a narrowband channel, the joint effect of wireless environment 
yields a linear model:
\begin{equation}
    \label{Eq: 2}
    y = h \cdot x + w
\end{equation}
where $x$ and $y$ are the transmitted and received signals, $h$ is the channel state information (CSI) and $w$ is the additive white Gaussian noise, all represented by complex values. Therefore, the channel response of subcarrier $i$  can be estimated by {\color{blue}$\hat{h}:=y/x$}. CSI characterizes the channel response with both the amplitude and the phase:
\begin{equation}
    |\hat{h}| = \frac{|y|}{|x|}
\end{equation}
and
\begin{equation}
    \angle{\hat{h}} = \angle{y} - \angle{x}.
\end{equation}

The availability of CSI in WiFi systems has fostered a plethora of applications including indoor localization and activity sensing. The former leverages an antenna array to collect the signals arriving at different antennas so that the Angle of arrival (AoA) and/or the time of flight (ToF) can be estimated for positioning \cite{xiong2013arraytrack}\cite{xiong2015tonetrack}. The latter takes the variation in the amplitude and phase 
of CSI over time as a feature that reflects human activities in a nearby wireless link.

\subsection{Weighted K-Nearest Neighbors}

We next describe weighted $k$-Nearest Neighbors (WKNN)\cite{5408784}, a statistical learning technique commonly used 
in WiFi indoor positioning systems as the matching rule. WKNN computes the distances (typically Euclidean distances) 
between each fingerprint in the database and a test sample, and picks up $k$ fingerprints with the smallest distances. 
Then, the estimation is made by taking the weighted average of the positions of the $k$ fingerprints. Those of smaller distances are endowed with relatively larger weights:
\begin{eqnarray}\label{eq: wknn}
\boldsymbol{\hat{p}} &= & \sum_{i = 1}^{k}w_i \ \boldsymbol{p}_i\\
w_i &= &\frac{1/\varepsilon_i}{\sum_{j=1}^k 1/\varepsilon_j}
\end{eqnarray}
\noindent
where $\boldsymbol{\hat{p}}$ is the estimated location, $\boldsymbol{p}_i$ is the position of the $i$th neighbor, and $\varepsilon_i$ is the corresponding distance.

\subsection{Transfer Learning}
We briefly describe the basic principle of transfer learning that is used in our system hereby. 
As is well known, machine learning is usually restricted by lack of sufficient data and fails to gain accurate knowledge. 
Transfer learning, a branch of machine learning, aims to apply the knowledge gained in solving one problem to address 
a different but related problem where limited or even no labeled data is available.
For instances, the knowledge acquired from learning to recognise cars can be applied to recognizing trucks, or 
the knowledge learned from CSI fingerprints in one location to constructing a prediction model in another location. 

In general, the concepts of a domain and a task are involved in transfer learning \cite{5288526}. 
A domain $\mathcal{D}$ consists of a feature space $\mathcal{X}$ and a marginal probability distribution 
$P(\boldsymbol{x})$ defined over this feature space where $\boldsymbol{x} = [x_1, x_2, \cdots, x_n] \in \mathcal{X}$. 
{\color{blue}A task $\mathcal{T}$} contains a label space $\mathcal{Y}$ and a conditional probability distribution 
$P(\boldsymbol{y}|\boldsymbol{x})$ that is a model learned from the training data. 
Transfer learning contains two domains: a source domain and a target domain denoted by $\mathcal{D}_s$ and $\mathcal{D}_t$ respectively. The domain $\mathcal{D}_s$ 
has the task $\mathcal{T}_s$ and the domain $\mathcal{D}_t$ has the task $\mathcal{T}_t$.\\ 
\indent
Given the adequate knowledge of the source domain $\mathcal{D}_s$ and a small amount of labeled data in the target domain $\mathcal{D}_t$, one popular method of transferring the information is to find a latent subspace for the source and target data so that the difference of probability distribution $P(\boldsymbol{x})$ and $P(\boldsymbol{y}|\boldsymbol{x})$ between the two domains are minimized. In the transfer learning, we adapt maximum mean discrepancy (MMD) \cite{NIPS2006_3110} as metrics, a non-parametric metric of the distribution difference. Given the samples from two domains $\{{\boldsymbol{x}}_s^i\}$ and $\{{\boldsymbol{x}}_t^i\}$, it computes the distance between averages of the sample projected into the subspace
\begin{equation}\label{eq: mmd}
    \mathrm{MMD} = {\left\| \frac{1}{n_s}\sum_{i=1}^{n_s}f(\boldsymbol{x}_s^i) - \frac{1}{n_t}\sum_{i=1}^{n_t}f(\boldsymbol{x}_t^i)  \right\|}^2
\end{equation}
where $f$ indicates the function that maps samples to the latent subspace, and ${n_s}$ and ${n_t}$ are the numbers of samples in the source and target domains. As shown in Equation (\ref{eq: mmd}), MMD equals zero when the two distributions in the subspace are the same. With the latent subspace found, a very limited amount of data at the target domain is capable of making classification with certainty.

%% file: 3_overview.tex
\section{System Overview} 
\label{sec:overview}

In this section, we articulate the advantages of CRISLoc over the state-of-the-art systems, followed by the description of system 
architecture. 

\begin{figure*}[t]
    \centering
	\setlength\abovecaptionskip{-1pt}
	\setlength\belowcaptionskip{-1pt}
	\begin{minipage}[t]{1\linewidth}
		\centering
		\subfloat[Intel 5300 CSI tool.]
		{\includegraphics[width=0.3\textwidth]{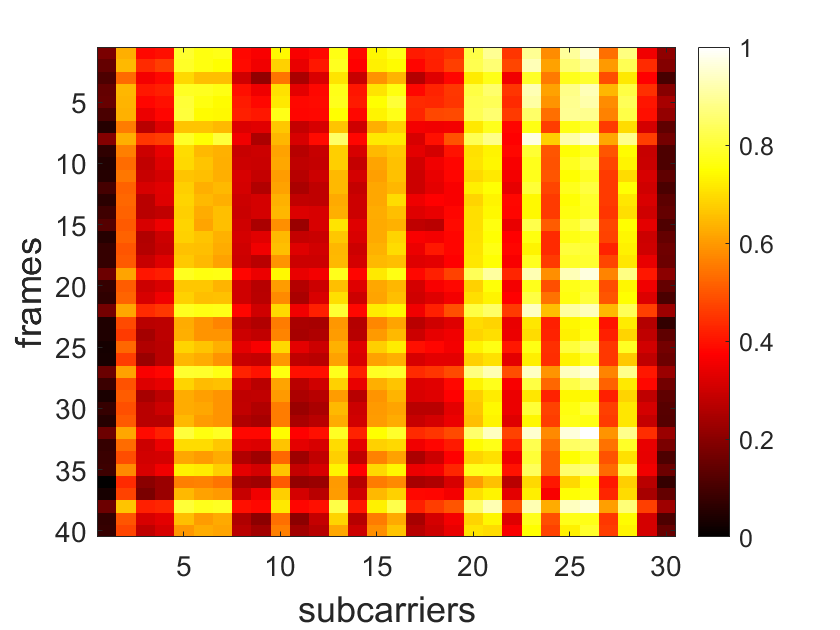}
		\label{fig: hotTool}}
		\hspace{0.05cm}
		\subfloat[Nexmon.]
		{\includegraphics[width=0.3\columnwidth]{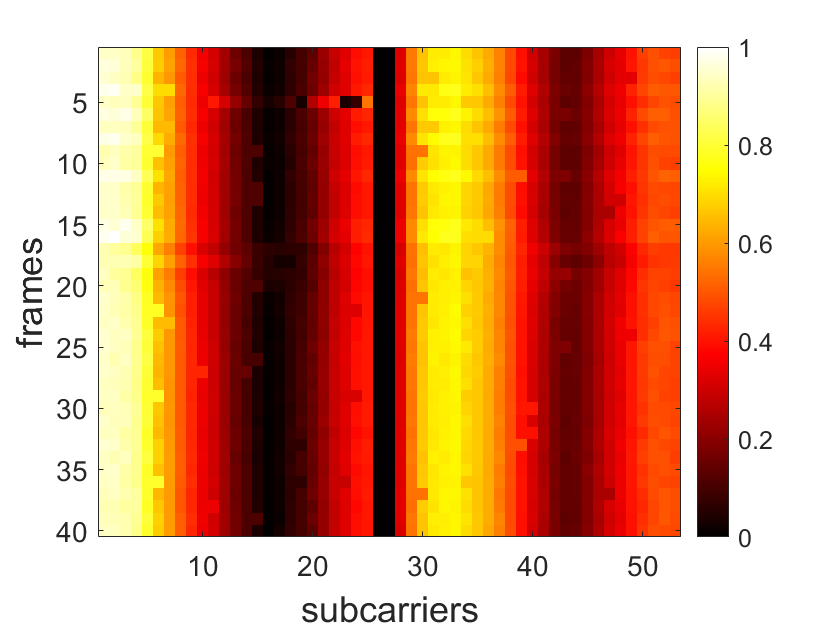}
		\label{fig: hot_nexmon}}
		\hspace{0.05cm}
		\subfloat[Variance Comparison.]
		{\includegraphics[width=0.3\textwidth]{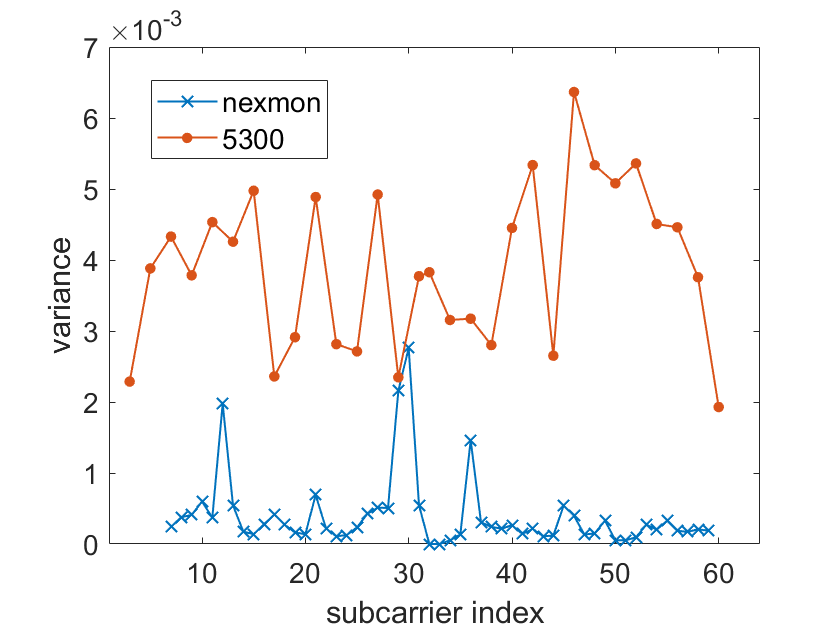}
		\label{fig: hotCompare}}
		\hspace{0.05cm}
		\caption{(a) and (b) are the normalized CSI amplitude of 40 continuous samples on all available subcarriers. (c) is the variance of each subcarrier over 40 samples.}
		\label{fig: hot}
	\end{minipage}
\end{figure*}

\subsection{Advantages of CRISLoc}

Intuitively, CRISLoc utilizes the CSI fingerprints of commodity smartphones for indoor localization. 
A crucial question is why the smartphone CSI is a better choice than the Intel 5300 CSI Tool  
and the smartphone RSS in fingerprinting. The reasons are summarized in two aspects.

\begin{itemize}
\item \textbf{Practicability.} i) Whenever a WiFi frame arrives at a receiver, CSI can be acquired by the unencrypted training sequence and the source AP can be identified by the unencrypted source MAC address\cite{7786995}, where public key and private key are not required, i.e., a successful connection is not required. It is rational for CRISLoc to overhear the transmitted frames in the air to acquire CSI, while 
Intel 5300 CSI Tool requires the successful connection to each AP at a time. 
When the surrounding APs are password protected or 
operated by a WLAN controller for roaming management, the smartphone can still be used for CSI indoor localization, but not 
the Intel 5300 CSI Tool; ii) Data frames, beacons and ACK frames can all be exploited for CSI acquisition in the smartphone, 
while only data fames are useful in the Intel 5300 CSI Tool. 
The above prominent properties not only make CSI fingerprinting ubiquitous, but also significantly reduce the time of 
conducting site survey.

\item \textbf{Performance.} i) The CSI of the smartphone is more stable than that of the Intel 5300 CSI Tool: as shown in Fig. \ref{fig: hot}, the CSI extracted by CSI tool fluctuates over time in $y$ axis; ii) 
The CSI on more subcarriers can be extracted from the smartphone than the CSI Tool, given the identical spectrum width. 
Therefore, the smartphone CSI has the potential to outperform Intel 5300 CSI Tool with more high quality fingerprints; 
iii) For CSI extracted from smartphones, there are 16 bits to represent the real and imaginary component in a subcarrier, while the number in other CSI tool is much less, i.e., 8 bits in Intel 5300 and 10 bits in Atheros. 
\end{itemize}

The augmentation of smartphone CSI brings new technical challenges spanning from the calibration of measured CSI to the detection of anomaly APs. The reconstruction of CSI-vector fingerprints when a fraction of APs are altered is especially difficult compared with that of simple RSS values.

\subsection{System Architecture}
\label{sec: architecture}

CRISLoc has two major objectives: one is to perform indoor localization with the collected 
CSI fingerprints; the other is to automatically 
detect the change of APs and reconstruct the fingerprinting database with minimal extra site surveys. 
The latter empowers the CSI fingerprinting based localization with anomaly detection, thus improving the positioning accuracy.

\begin{figure*}[ht]
\centering
\includegraphics[width=0.8\textwidth]{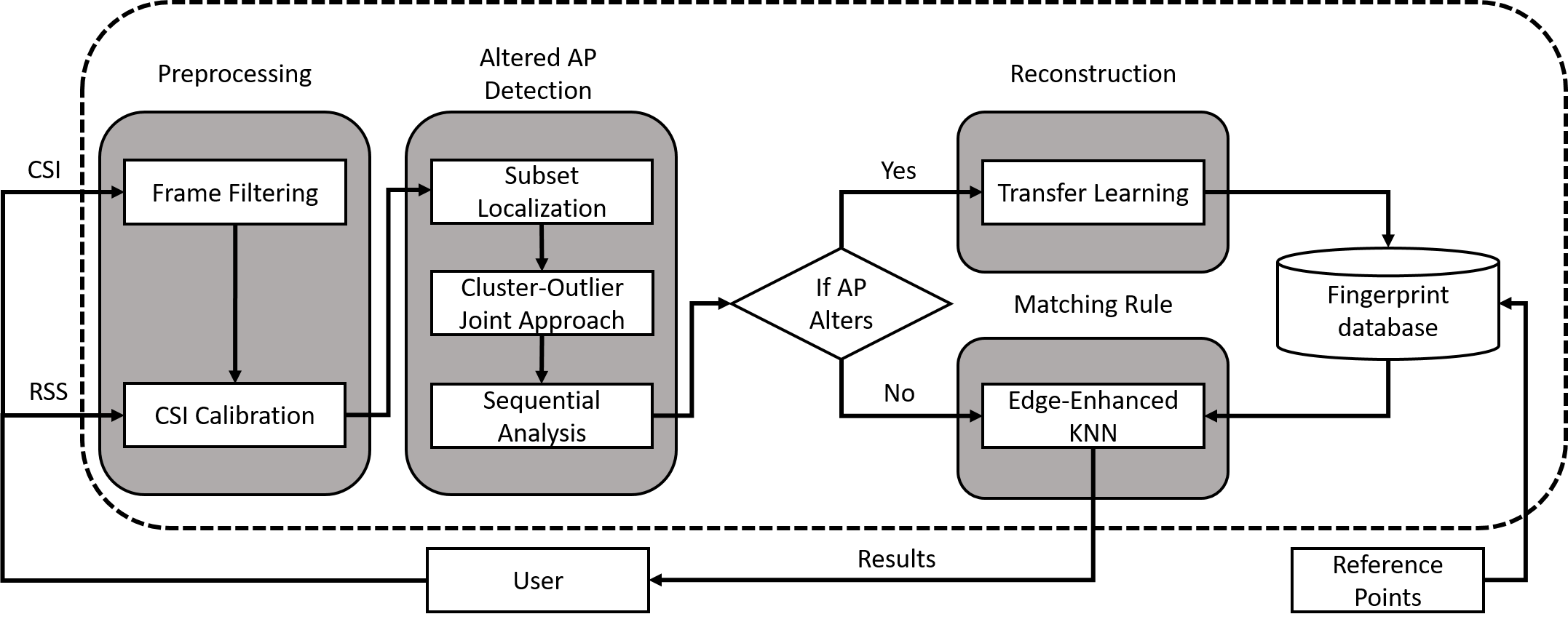}
\caption{System Architecture.}
\label{fig: systemOverview}
\end{figure*}
\indent

The overall architecture of CRISLoc is shown in Fig. \ref{fig: systemOverview} which consists of the pre-processing module, 
the altered AP detection module, the reconstruction module and the localization module. 

\textbf{Pre-processing.} The raw CSI data collected by a smartphone cannot be directly used to create the fingerprinting 
database because the AGC function distorts the measured amplitude of the received signal. 
Meanwhile, the frame filtering is adopted to remove 
abnormal frames. 

\textbf{Altered-AP Detection.} An obstacle preventing the usage of WiFi fingerprinting localization is the 
change of AP deployment. Once a fraction of APs malfunction or are moved to other locations, the fingerprint database 
becomes obsoleted. CRISLoc tackles this challenge by altered AP detection, including subset localization, cluster-outlier joint approach and sequential analysis. 

\textbf{Fingerprint Reconstruction.} CRISLoc adopts domain adaptation transfer learning to reconstruct 
the fingerprints of altered APs. After projecting the outdated fingerprints and the newly collected fingerprints of reference points (RPs)
onto a subspace, CRISLoc generates a new fingerprint database through minimizing their 
Euclidian distance. {\color{blue}Note that RP is a mobile terminal that continuously collect fingerprints at a certain position.}

\textbf{Matching Rule.} CRISLoc developes an edge enhanced $k$-nearest neighbours (EEKNN) approach to pinpoint 
a smartphone. EEKNN is capable of handling the corner scenarios where the previous approaches are prone to large errors.

%% file: 4_1_preprocessing.tex
\section{CSI Filtering and Anomaly Detection}
\label{sec: systemdesign}

In this section, we present filtering methods of smartphone CSI and 
algorithms to detect the set of altered APs. 

\subsection{Pre-processing}

\begin{algorithm}[htbp]
\caption{CSI Pre-processing}
\label{alg: Pre-processing}
{\bf Input: } raw CSI $\boldsymbol{x}^{i,j}$, RSSI $\boldsymbol{r}^{i,j}$\\
{\bf Output: } fingerprinting CSI $\boldsymbol{x}^{i,j}$ 
\begin{algorithmic}[1]

\For{each training/testing points $i$}
    \For{each CSI sample $\boldsymbol{x}^{i,j}$}
        \State{get the corresponding $\boldsymbol{r}^{i,j}$}
        \State{calculate $s$ according to Eq.\ref{eq: coefficient}}
        \State{$\boldsymbol{x}^{i,j} = \boldsymbol{x}^{i,j} / s$}
    \EndFor
    \State{calculate the mean $\boldsymbol{\mu}^i$ at point $i$}
    \State{calculate the covariance matrix $\mathbf{\Sigma}^i$ at point $i$}
    \For{each CSI sample $\boldsymbol{x}^{i,j}$}
        \State{$d(\boldsymbol{x}^{i,j}) = \sqrt{(\boldsymbol{x}^{i,j} - \boldsymbol{\mu}^i)^T\mathbf{\Sigma}^{i-1}(\boldsymbol{x}^{i,j} - \boldsymbol{\mu}^i)}$}
    \EndFor
    \State{sort($d(\boldsymbol{x}^{i,j})$)}
    \State{delete $\boldsymbol{x}^{i,j}$ with the highest 5\% $d(\boldsymbol{x}^{i,j})$}
\EndFor

\end{algorithmic}
\end{algorithm}

We only make use of the CSI amplitude and leave the CSI phase unexploited. The reasons are two-folded\cite{wang2015phasefi}. Firstly, the extracted CSI phase is not a fixed value. Due to carrier frequency offset (CFO) and sampling frequency offset (SFO), the estimation error of CSI phase occurs and accumulates over time. Besides, an extremely subtle change in the internal circuits (e.g. oscillators, phase lockers and amplifiers) may cause a remarkable drift in the phase estimate, making the estimated phase obsolete in a short period. Secondly, the feature of phases in a narrowband channel is less indicative than that of amplitude envelopes. If the phase estimates are expanded along the subcarriers, a linear correlation is observed due to the evenly spaced frequencies of different subcarriers. Such a simple feature is insufficient to serve the ``identity'' of a location. 

The CSI data is pre-processed by taking the following two steps: frame filtering, and CSI calibration {\color{blue}which are summarized in Algorithm \ref{alg: Pre-processing}.}

\indent
\subsubsection{Frame Filtering}
\begin{figure}[t]
	\centering
	\setlength\abovecaptionskip{-1pt}
	\setlength\belowcaptionskip{-1pt}
	\begin{minipage}[t]{1\linewidth}
		\centering
		\subfloat[abnormal CSI frames.]
		{\includegraphics[width=0.47\textwidth]{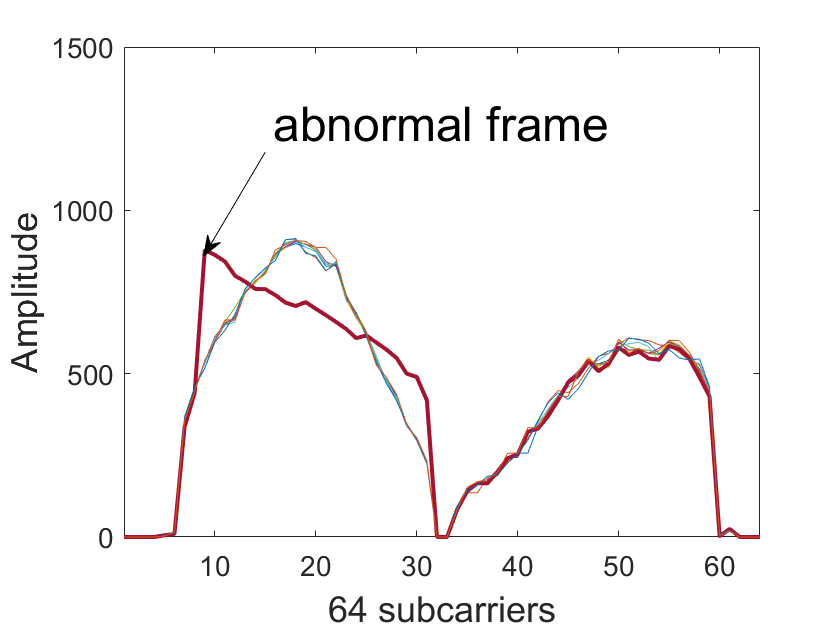}\label{fig: CSIOutlier}}
		\hspace{0.05cm}
		\subfloat[Mahalanobis distances.]
		{\includegraphics[width=0.47\columnwidth]{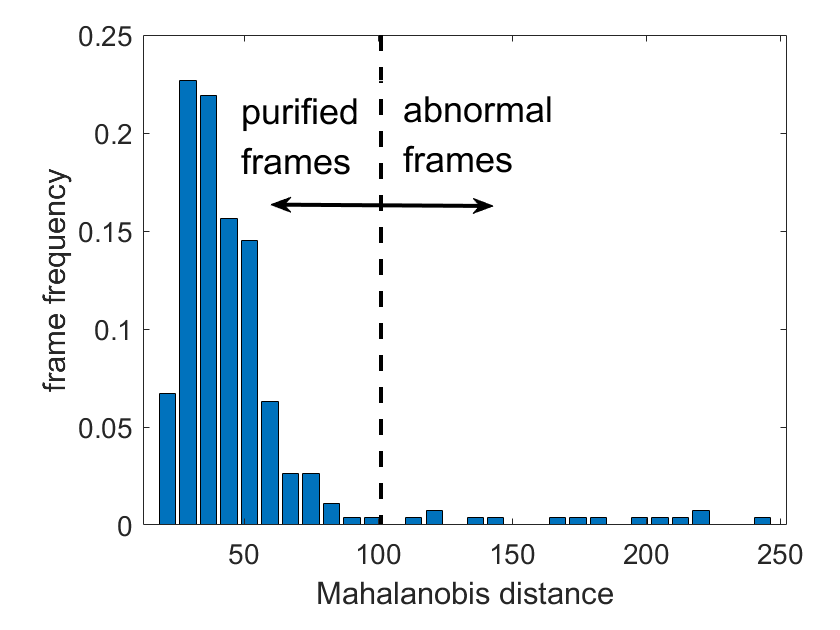}
		\label{fig: mdisHist}}
		\hspace{0.05cm}
		\caption{Mahalonobis distance collected under a certain circumstance, whose envelopes are shown in (a) and frequency histogram is shown in (b).}
	\end{minipage}
\end{figure}

Abnormal CSI measurement (one out of fifty frames), shown in Fig.\ref{fig: CSIOutlier}, may appear due to the environment noise. 
We develop a frame filtering approach based on Mahalanobis distance. In general, the Mahalanobis distance is defined as:
\begin{equation}\label{eq: m-distance}
d(\boldsymbol{x}) = \sqrt{(\boldsymbol{x} - \boldsymbol{\mu})^T\mathbf{\Sigma}^{-1}(\boldsymbol{x} - \boldsymbol{\mu})}
\end{equation}
where $\boldsymbol{x}$ is the sample vector, $\boldsymbol{\mu}$ is the arithmetic mean vector of a set of observations, and $\mathbf{\Sigma}$ is the covariance matrix. Compared to Euclidean distance, Mahalanobis distance is a metric that measures how an observation $\boldsymbol{x}$ is away from a given distribution, taking the covariance of elements in a vector into account. Note that a minimum number of CSI measurements are required to calculate the covariance matrix $\mathbf{\Sigma}$. Therefore, we recollect CSIs when the sampled frames are not enough.\\

\indent
The frame filter calculates the Mahalanobis distance of each frame 
received at the same point (site survey) and in a short period (user 
request). We evaluate the Mahalanobis distances of three hundred 
frames and plot them in a histogram (Fig. \ref{fig: mdisHist}). 
Here, x-coordinate represents the discretized interval of 
Mahalanobis distance and y-coordinate represents the fraction of 
data frames falling in a specific interval. 
Our experimental results show that the abnormal frames appears to be totally different from other frames and are prone to 
having so large Mahalanobis distances that an obvious gap appears between the most frames and the abnormal frames. By setting an adaptive threshold located on such gap
that filters out 5\% of frames with large Mahalanobis distances, 
we obtain the purified CSI frames suitable for fingerprinting.\\

\begin{figure}[t]
\vspace{-0.5cm}
	\centering
	\setlength\abovecaptionskip{-1pt}
	\setlength\belowcaptionskip{-1pt}
	\begin{minipage}[t]{1\linewidth}
		\centering
		\subfloat[Uncalibrated CSI]
		{\includegraphics[width=0.45\textwidth]{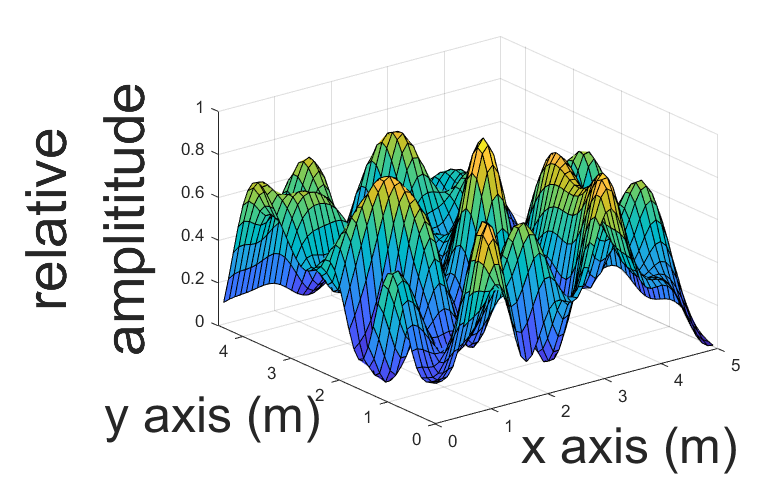}\label{fig: CSI_ori}}
		\hspace{0.05cm}
		\subfloat[Calibrated CSI]
		{\includegraphics[width=0.45\textwidth]{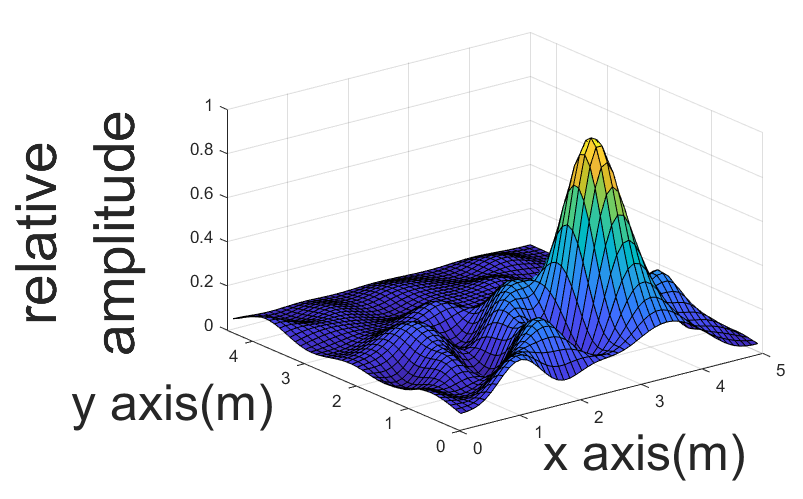}\label{fig: CSI_scaled}}
		\hspace{0.05cm}
		\caption{The radio map of the uncalibrated CSI (a) and calibrated CSI (b). The AP locates at $(3.5,2)$.}
		\label{fig: path loss}
	\end{minipage}
\end{figure}

\subsubsection{CSI Calibration} 

The CSI data extracted from smartphones cannot be directly used due 
to that the AGC scheme at the receiver 
magnifies the amplitude of the original CSI. 
This is to say, the extracted CSI is multiplied by an unknown factor whose value changes as the user moves.
As a result, the power of CSI no longer follows the basic  
path loss principle and the basic rationale of fingerprinting based localization fails 
miserably. \\ 
\indent
To solve this problem, we propose to rescale the extracted CSI so that AGC is canceled using RSS based on the fact that RSS is obtained before AGC while CSI is obtained after AGC\cite{7786995}\cite{Halperin_csitool}. 
Since AGC is a linear time-invariant (LTI) system and is homogenious for every subcarrier, the ratio between the CSI of a couple of different subcarriers remains the same. Given that the sum of CSI squared over all the subcarriers should be consistent with RSS, we multiply the extracted CSI of all subcarriers by a single coefficient:
\begin{equation}
    \mathit{s} =  \sqrt{\frac{10^{RSS/10}}{\sum{CSI_i^2}}},
    \label{eq: coefficient}
\end{equation}
\noindent
which yields the CSI before AGC. Here, $CSI_i$ is the extracted CSI of the 
$i^{th}$ subcarrier and $RSS$ is the received signal quality in dBm{\color{blue}, which should be modified to mW}. In this way, the power of the rescaled CSI equals to the corresponding 
RSS and AGC is thus cancelled. \\
\indent
The effectivenss of CSI calibration is demonstrated in Fig. \ref{fig: path loss}, where we plot the relative CSI amplitude on one subcarrier (shown in the z axis) in a 4.5 meters by 5 meters area for both uncalibrated and calibrated CSI. The origin of the figure is one corner of the area and the AP locates at (3.5, 2). It is easy to see that the uncalibrated CSI does not follow the signal path loss, while the calibrated CSI does: a smartphone closer to the AP usually collects a CSI with a larger amplitude.

%% file: 4_2_detection.tex
\subsection{Altered AP Detection}

The biggest challenge of fingerprinting based 
localization is the sensitivity to environmental changes, especially the relocation of APs. 
One crucial question arises: \textit{can we 
accurately detect the change of APs and 
revamp the fingerprint database accordingly?
}

\subsubsection{Subset Localization} 

\indent
The basic idea of altered AP detection is to discover the discrepancy 
of localization results using different subsets of available APs.
We can acquire different estimations of the user position with multiple AP subsets. By scrutinizing these results, we observe that
the subsets containing one or more altered APs yield scattered positions, while those 
without altered APs have agminated positions. 
We hence detect the altered APs by distinguishing these two kinds of subsets.

\indent
The setting of the number of APs in each subset is crucial. Let $\mathbf{P}$ be the whole set of APs, and $P_s$ be a subset of $\mathbf{P}$. When $P_s$ contains a very few 
APs, $P_s$ tends to have large localization errors even if 
it does not contain 
altered APs. On the contrary, $P_s$ is likely to perform very well if there are 
a lot of well-functioning APs but few altered APs. Therefore, a minimum and 
a maximum values are configured to select 
the subsets of APs (minimum is three and maximum is five in our implementation).
For each subset, we estimate the location of 
a smartphone using the matching rule
in Section \ref{sec: localization}. 
When the APs are densely deployed, we define a $k$-neighborhood for each AP, where there are only $k$ spatially nearest APs, excluding itself, in its $k$-neighborhood. For each $k$-neighborhood, we select three to five APs in it as subsets. 
We tend to choose the $k$ as great as the computation supports and time permits. Meanwhile, it is guaranteed that the frequency that each AP appears in the these subsets with the same neighborhood are the same.
We only select as few of the $k$-neighborhoods as they contain all the APs, and examine them by the following approach respectively.

\begin{figure}[t]
    \centering
    \includegraphics[width=0.6\columnwidth]{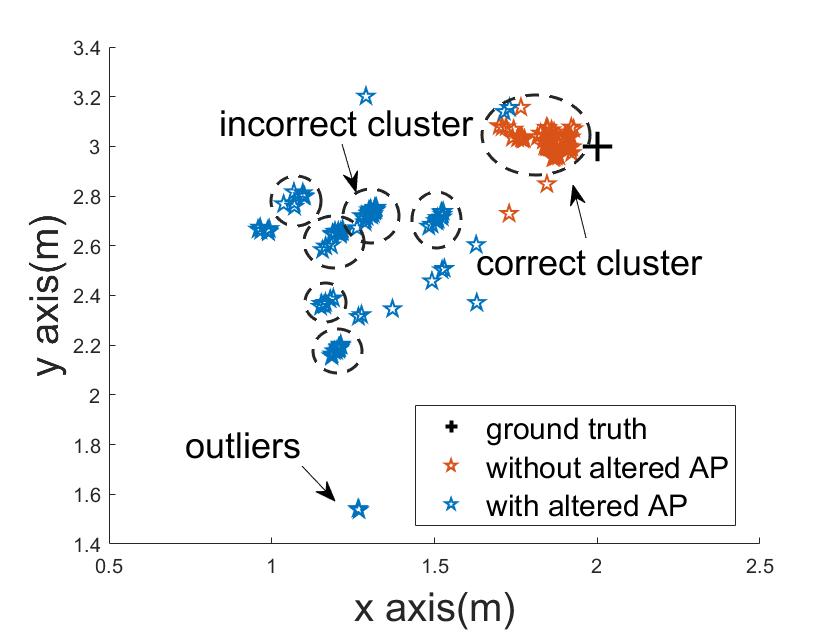}
    \caption{Localization results with multiple subsets}
    \label{fig: subsets}
\end{figure}

\subsubsection{Cluster-Outlier Joint Approach}
\indent
Adopting the joint approach is crucial: a clustering method 
may cause an overestimation of altered APs; 
{\color{blue}an outlier} detection approach, on the 
contrary, may underestimate the number of 
altered APs. The joint cluster-outlier approach has the potential to achieve both 
high precision and high recall of detection.

Intuitively, as shown in the Fig.\ref{fig: subsets}, the localization results coming 
from the subsets without altered APs 
tend to be in the same and the 
largest cluster and close to the ground truth. The largest cluster is deemed as the ``ground truth'' cluster (GTC) of localization mostly containing unaltered APs. Similarly, we deem those localization results that is not included in the GTC as non-GTC points and consider all the non-GTC points as the localization results of the subsets containing altered APs. In order to detect altered APs, we examine the non-GTC points.
However, due to the estimation errors, 
the localization results may be grouped into 
multiple clusters that are of comparable 
sizes. As shown in Fig. \ref{fig: subsets}, 
our experiments demonstrate the existence 
of several clusters that are hard to 
differentiate their relative significance. 
If we wrongly pick the ``largest'' cluster, 
the detection of altered AP will fail miserably.
After the good APs being removed and the altered
APs being kept, the accuracy of localization is 
worse off than that with the original 
CSI fingerprints. To avoid such kinds of mistakes, 
we only examine the outliers of localization results under such circumstances. 
Fig. \ref{fig: subsets} shows that the marjority of the outliers are generated by the subsets with altered APs. On the other hand, the localization results 
of the subsets of unaltered APs are inclined to be classified 
together, and may not give rise to many outliers.

The cluster-outlier joint approach operates in three steps. 
First, the clustering algorithm runs in which the classical 
DBSCAN method \cite{Ester:1996:DAD:3001460.3001507} is employed. 
Denote by $C_{1^{st}}$ 
and $C_{2^{nd}}$ the sizes of the largest and the second largest 
clusters.

\indent
Second, given $\frac{C_{1^{st}}}{C_{2^{nd}}}$ and a threshold $r_0$, we count the total frequency $f_i$ that the $i$th AP appears as the non-GTC points when the ratio 
 is greater than a certain threshold $r_0$. Otherwise, we transit to the second step by exploiting 
the outliers. Those APs with higher frequency of occurrence are more likely to alter. 

Third, with the summed-up frequency of each AP, we employ Jenks natural breaks classification method \cite{Jenks} to classify the altered and unaltered APs. Note that Jenks method classifies symmetrically, which disobeys the truth that the appearance of altered APs and unaltered APs are asymmetric. The variance of frequencies of altered APs are multiplied by an adaptive factor $\eta$. The larger $\eta$ is, the fewer number of altered APs the system is likely to claim. Then, the classification $\mathbf{P} = \{P_1, P_2\}$ can be found by minimizing the weighted sum of the squared deviations from the class means:

\begin{equation}
    \mathbf{P} = \arg\min\limits_{\mathbf{P}}\Big(\sum_{i \in P_{1}}{(f_i - \bar{f}_{P1})}^2 + \eta \sum_{i \in P_{2}}{(f_i - \bar{f}_{P2})}^2\Big).\\
\end{equation}
The whole algorithm of cluster-outlier joint approach is shown in Algorithm \ref{alg: joint approach}.

\begin{algorithm}[t]
\caption{Cluster-Outlier Joint Approach}
\label{alg: joint approach}
{\bf Input: }localization results with AP subsets\\
{\bf Parameter: }$r_0$\\
{\bf Output: atered AP}
\begin {algorithmic}[1]
\State {perform DBSCAN}
\If{$\frac{|C_{largest}|}{|C_{second\ largest}|} \leq r_0$}
\State {count $f_i$ of the $i$th AP at non-GTC points}
\Else
\State {count $f_i$ of the $i$th AP in outliers}
\EndIf
\State {sort $f_i$}
\State {perform Jenks method on $f_i$}
\end{algorithmic}
\end{algorithm}

\subsubsection{Sequential Analysis} 
Sequential analysis is adopted after the joint approach to improve the accuracy of the detection. The basic idea is to combine several samples to come out a more reliable result. In addition, sequential analysis solves an inevitable problem joint approach incurs: Jenks method always classifies into two classes, both of which contains at least one element. Therefore, {\color{blue}$P_2$} contains at least one AP even if there's no AP alters and thus false alarm occurs.\\
\indent
In sequential analysis, we first take $min\_seq$ number of consecutive samples into account, and the ratio of alarms that each AP change is calculated. The ratio is taken as the reliability level $l$ of asserting an AP as altered. On the one hand, if $l$ is higher than a reliability level threshold $l_0$, an AP can be claimed as altered with certainty. {On the other hand, if the ratio is lower than $1-l_0$, the AP is claimed as unaltered.} Otherwise, a judgment whether an AP is altered can not be made for sure, in which more samples are taken into account until the judgement can be made. Note that this process will be interrupted with an explicit output by comparing to 0.5 if the number of the samples we check exceeds an upper bound $max\_seq$. The procedure of detecting altered AP is summarized in Algorithm \ref{alg: sequential analysis}.
As for the cases that no AP alters, the AP which appears most frequently as outliers or non-GTC points differs for different test samples. That is to say, no AP is frequently classified into the altered class. \\

\begin{algorithm}[t]
\caption{Sequantial Analysis}
\label{alg: sequential analysis}
{\bf Input: }results of Cluster-Outlier joint approach\\
{\bf Parameter: }$min\_seq$, $max\_seq$, $l_0$\\
{\bf Output: } whether AP alters
\begin{algorithmic}[1]
\State {$l = 0$, $n\_seq$ = $min\_seq$}
\For {$min\_seq$ samples}
\State {Perform Cluster-Outlier Joint Approach}
\State {update $l$}
\EndFor
\While {$1-l_0 < l$ \textbf{and} $l < l_0$ \textbf{and} $n\_seq \leq max\_seq$}
\State {Perform Cluster-Outlier Joint Approach}
\State {update $l$}
\State {$n\_seq = n\_seq + 1$}
\EndWhile
\If {$n\_seq \leq max\_seq$}
\If {$1-l_0 > l$}
\State \Return{AP is unaltered}
\Else
\State \Return{AP is altered}
\EndIf
\Else
\If {$l < 0.5$}
\State \Return{AP is unaltered}
\Else
\State \Return{AP is altered}
\EndIf
\EndIf
\end{algorithmic}
\end{algorithm}

%% file: 4_3_transfer.tex
\section{CSI Reconstruction and Localization}
\label{sec:transfer}

In this section, we propose a novel transfer learning method to 
reconstruct the CSI fingerprints based the maximum mean discrepancy measure. An edge-enhanced CSI matching rule is designed to perform indoor 
localization.

\subsection{Fingerprint Reconstruction}

When APs are altered, the CSI fingerprints seem to be useless. Yet, recollecting new CSI is time-consuming and economically inefficient. An interesting question is 
whether the outdated CSI fingerprints combined with 
a few fresh CSI samples from the reference points can 
be used to generate the updated fingerprints without 
cumbersome survey of all the sites. A reference point (RP) is a mobile terminal that continuously collect the fingerprints such as CSI from surrounding APs. 
We observe that the factors influencing CSI such 
as the building layout usually change very gently despite the change of the 
location of an AP. The path loss pattern of spatially 
adjacent points may change in a similar way. 
Therefore, we leverage the transfer learning approach 
to reconstruct the CSI database at new scenarios 
with the knowledge gained from the outdated fingerprints. 

{\color{blue}Similar to \cite{6751384},} the main target of transfer learning is to find a transform matrix $\mathbf{A}$ that projects both the outdated CSI data, serving as the source domain, and the updated CSI data, serving as the target domain, into a subspace where data distribution is matched. The properties of CSI radio map should be preserved at the same time. By projecting outdated CSI data into the subspace, we reconstruct the CSI fingerprints which achieve high localization accuracy. It is assumed that training points include all RPs. In the following, we present how to find the optimal transform matrix $\mathbf{A}$. The notations that we use are listed in Table \ref{table: notation1}.

\begin{table}[htbp]
\caption{Notations Used in Fingerprint Reconstruction}
\label{table: notation1}
\begin{center}
\renewcommand\arraystretch{1.2}
\begin{tabular}{|c|c|}
\hline
Notation & Definition           \\ \hline
$\boldsymbol{1}$ & All-one column vector\\
$\mathbf{I}$ & Identity matrix\\
$m$ & The dimension of one AP's CSI samples \\
$M$ & The dimension of subspace \\
$n_s^i$ & The number of the outdated samples at point $i$\\
$n_t^i$ & The number of the updated samples at point $i$\\
$c_s$ & The number of points with outdated labeled samples\\
$c_t$ & The number of points with updated labeled samples\\
$\mathbf{A}$         & Transform matrix, $\mathbb{R}^{m\times{M}}$\\
$\boldsymbol{x}_s^{i, j}$ &The $j$th outdated CSI sample at point $i$, $\mathbb{R}^{m\times{1}}$\\
$\boldsymbol{x}_t^{i, j}$ &The $j$th updated CSI sample at point $i$, $\mathbb{R}^{m\times{1}}$\\
$\bar{\boldsymbol{x}}_s^i$  & The average of $\boldsymbol{x}_s^{i, j}$ over samples $j$ at point $i$\\
$\bar{\boldsymbol{x}}_t^i$  & The average of $\boldsymbol{x}_t^{i, j}$ over samples $j$ at point $i$\\
$\bar{\boldsymbol{x}}_s$ & The average of the outdated samples over points $i$\\
$\mathbf{X}_s^i$ & $[\boldsymbol{x}_s^{i,1}, \boldsymbol{x}_s^{i,2}, \cdots, \boldsymbol{x}_s^{i,n_i}]\in\mathbb{R}^{m\times{n_s^i}}$\\
$\mathbf{X}$ & $[\mathbf{X}_s^1\ \mathbf{X}_s^2 \cdots \mathbf{X}_s^{c_t}\ \mathbf{X}_t^1 \cdots  \mathbf{X}_t^{c_t}]$\\
$N_i$ & The set of neighbours of point $i$\\
$F_i$ & The set of non-neighbours of point $i$\\
\hline
\end{tabular}
\end{center}
\end{table}

\subsubsection{Minimizing Distance between Distributions} 
\label{sec: disDomain}
\indent
The auto-update of fingerprints is required to align the outdated CSI data with the up-to-date WiFi environment. 
In our task of fingerprint reconstrction, the source domain is the outdated CSI data and the target domain is the newly collected CSI data. Given that the position of the AP changes, we are likely to collect different CSI data at the same position. That is to say, the distribution $P(\boldsymbol{x}, \boldsymbol{y})$ changes, where $\boldsymbol{x}$ is the CSI vectors and $\boldsymbol{y}$ is the position where data is collected. In order to minimize the difference of distribution between source and target domains, we use the revised maximum mean discrepancy (MMD) measure.
A variable $s_t$ is used to measure the extent to which two distributions resemble one another according to MMD. $s_t$ is measured by summing up the distances between the means of the projected outdated samples and updated ones over the points where both updated labeled samples and outdated samples exist. Here we use the Euclidean distance and rewrite the Euclidean distance using matrix traces.

\begin{equation}
\small
\begin{aligned}
s_t&=\sum_{i=1}^{c_t}||\mathbf{A}^T \bar{\boldsymbol{x}}_s^i - \mathbf{A}^T\bar{ \boldsymbol{x}}_t^i||^2\\
&=\sum_{i=1}^{c_t}\mathrm{Tr}\big((\mathbf{A}^T \bar{\boldsymbol{x}}_s^i - \mathbf{A}^T\bar{ \boldsymbol{x}}_t^i)(\mathbf{A}^T \bar{\boldsymbol{x}}_s^i - \mathbf{A}^T\bar{ \boldsymbol{x}}_t^i)^T\big)\\
&=\sum_{i=1}^{c_t}\mathrm{Tr}\big(\mathbf{A}^T
[\;\mathbf{X}_s^i \ \mathbf{X}_t^i \;]
{\left[\begin{array}{cc}
\frac{\boldsymbol{1}\boldsymbol{1}^T}{(n^i_s)^2} &  -\frac{\boldsymbol{1}\boldsymbol{1}^T}{n^i_sn^i_t}\\
\frac{\boldsymbol{1}\boldsymbol{1}^T}{n^i_sn^i_t} &  -\frac{\boldsymbol{1}\boldsymbol{1}^T}{(n^i_t)^2}
\end{array} 
\right ]}
{\left[\begin{array}{c}
{\mathbf{X}_s^i}^T\\
{\mathbf{X}_t^i}^T 
\end{array} 
\right ]})
\mathbf{A}\big)\\
&=\sum_{i=1}^{c_t}\mathrm{Tr} (\mathbf{A}^T\mathbf{X}\mathbf{M}_i\mathbf{X}^T\mathbf{A})\\
&=\mathrm{Tr}(\mathbf{A}^T\mathbf{X}\mathbf{M}\mathbf{X}^T\mathbf{A}),\\
\end{aligned}
\label{eq:s_t}
\end{equation}

where $\mathbf{M}_i$ and $\mathbf{M}$ are defined as:
\begin{align}
(\mathbf{M}_i)_{pq}&=\left\{
\begin{array}{rcl}
\frac{1}{(n^i_s)^2}  & & \boldsymbol{x}_p, \boldsymbol{x}_q\in\mathbf{X}_s^i\\
\frac{1}{(n^i_t)^2} & &  \boldsymbol{x}_p, \boldsymbol{x}_q\in\mathbf{X}_t^i\\
-\frac{1}{n^i_sn^i_t} & & \left\{
\begin{array}{l}
\boldsymbol{x}_p\in\mathbf{X}_s^i,  \boldsymbol{x}_q\in\mathbf{X}_t^i\\
\boldsymbol{x}_q\in\mathbf{X}_s^i,  \boldsymbol{x}_p\in\mathbf{X}_t^i
\end{array}\right.\\
0 & & otherwise
\end{array}\right.\\
\mathbf{M}&=\sum_{i=1}^{c_t}\mathbf{M}_i.
\label{eq: tf-Mi}
\end{align}
\normalsize

\subsubsection{Minimizing Intra-Class Distance}
\label{sec: disIntra}
When minimizing distribution distances, important properties of CSI data such as stability should be maintained. One aspect of data properties is that the projected samples within the same class ought to be as clustered as possible. Intra-class distance $s_i$ measures the dispersion of the samples collected at point $i$:
\begin{equation}
\begin{aligned}
s_i &= \sum_{j=1}^{n_s^i}\mathrm{Tr}\big(\mathbf{A}^T(\boldsymbol{x}_s^{i, j} - \bar{\boldsymbol{x}}_s^i)(\boldsymbol{x}_s^{i, j} - \bar{\boldsymbol{x}}_s^i)^T\mathbf{A}\big)\\
& = \mathrm{Tr}\big(\mathbf{A}^T\mathbf{X}_s^i(\mathbf{I} - \frac{1}{n_s^i}{\boldsymbol{1}\boldsymbol{1}^T})(\mathbf{X}_s^i)^T\mathbf{A}\big).
\end{aligned}
\label{eq: tf-intra}
\end{equation}
\indent
Equation \eqref{eq: tf-intra} measures the dispersion within one class. Summing up $s_i$ of all classes yields the intra-class distance of all samples $s_w$:
\begin{equation}
s_w =\mathrm{Tr}(\mathbf{A}^T\mathbf{P}_s\mathbf{A}), \ \mathbf{P}_s= \sum_{i=1}^{c_s}\mathbf{X}_s^i(\mathbf{I} - \frac{1}{n_s^i}{\boldsymbol{1}\boldsymbol{1}^T})(\mathbf{X}_s^i)^T.
\label{eq:s_w}
\end{equation}

\subsubsection{Maximizing Inter-class Distance}
\label{sec: disInter}
In order to distinguish classes more readily, the separation between classes should be maximized. The separation $s_b$ is measured by the global dispersion $s_g$ minus the intra-class distance $s_w$.
The global dispersion $s_g$ is defined as summing up all the distances between every projected outdated sample and $\mathbf{A}^T\bar{\boldsymbol{x}}_s$, the average of projected outdated samples over all site-survey points.
\begin{equation}
s_g = \sum_{i=1}^{c_s}\sum_{j=1}^{n_s^i}\mathrm{Tr}\big(\mathbf{A}^T(\boldsymbol{x}_s^{i, j} - \bar{\boldsymbol{x}}_s)(\boldsymbol{x}_s^{i, j} - \bar{\boldsymbol{x}}_s)^T\mathbf{A}\big)
\end{equation}
\noindent
Then we have
\begin{equation}
\label{eq:s_b}
\begin{aligned}
s_b =& s_g - s_w\\
=&\sum_{i=1}^{c_s}\sum_{j=1}^{n_s^i}\mathrm{Tr}\Big(\mathbf{A}^T\big(-\boldsymbol{x}_s^{i, j}\bar{\boldsymbol{x}}_s^T - \bar{\boldsymbol{x}}_s(\boldsymbol{x}_s^{i, j})^T + \bar{\boldsymbol{x}}_s\bar{\boldsymbol{x}}_s^T + \\&\ \boldsymbol{x}_s^{i, j}(\bar{\boldsymbol{x}}_s^i)^T + \bar{\boldsymbol{x}}_s^i(\boldsymbol{x}_s^{i, j})^T -\bar{\boldsymbol{x}}_s^i\bar{\boldsymbol{x}}_s^i\big)^T\mathbf{A}\Big)\\
=&\mathrm{Tr}(\mathbf{A}^T\mathbf{Q}_s\mathbf{A}), \ \mathbf{Q}_s= \sum_{i=1}^{c_s}n_s^i(\bar{\boldsymbol{x}}_s^i - \bar{\boldsymbol{x}}_s)(\bar{\boldsymbol{x}}_s^i - \bar{\boldsymbol{x}}_s)^T.
\end{aligned}
\end{equation}

\subsubsection{Closer Points Sharing Similar CSI Data}
\begin{figure}[t]
    \centering
    \includegraphics[width=0.5\columnwidth]{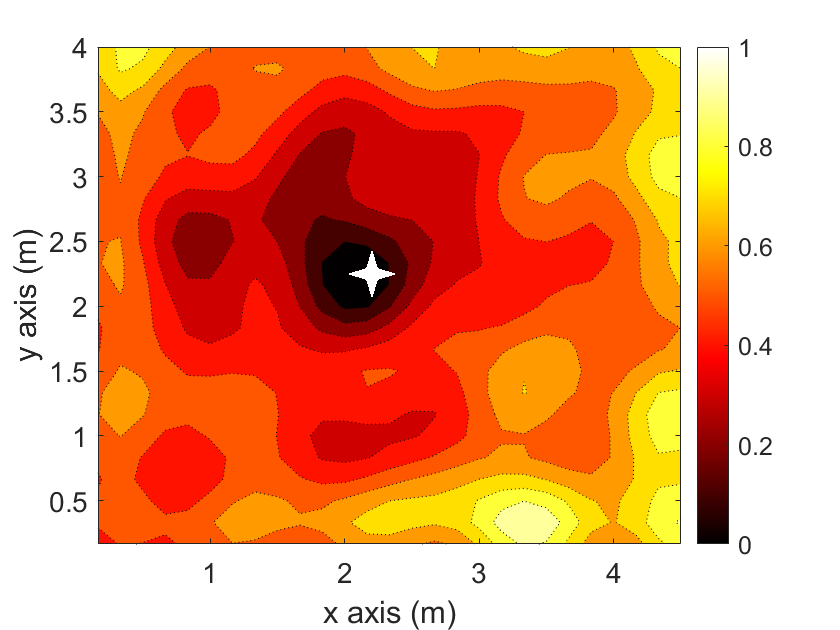}
    \caption{Inter-class distance of all points with respect to the point $x = (2.25m, 2.25m)$. One corner of the room is chosen as the origin $(0,0)$.}
    \label{fig: neighbourDis}
\end{figure}

One special data property of CSI radio map is that the points spatially closer to one other share similar CSI fingerprints, as shown in Fig. \ref{fig: neighbourDis}. Generally, the nearby points tend to have an much lower distance in fingerprints compared with points far apart. If the rule does not hold true, the system may pick up `CSI neighbors' spatially far away from the user as its predictions. Therefore, we set a rule that the projected CSI distances between neighbors $d_N$ should be smaller than the distances between far away points $d_F$.
\begin{equation}
    d_N = \sum_{i=1}^{c_s}\frac{1}{|N_i|}\sum_{k\in{N_i}}{\mathrm{Tr}\big((\mathbf{A}^T\bar{\boldsymbol{x}}_s^i - \mathbf{A}^T\bar{\boldsymbol{x}}_s^k)(\mathbf{A}^T\bar{\boldsymbol{x}}_s^i - \mathbf{A}^T\bar{\boldsymbol{x}}_s^k)^T\big)}
\end{equation}
where $N_i = \{k|k\ \mathrm{is\ the\ spatial\ neighbor\ of}\ i\}$. Let $\Delta \boldsymbol{x}_{N}^k = \bar{\boldsymbol{x}}_s^i - \bar{\boldsymbol{x}}_s^k$ for $k \in N_i $, we have 
\begin{equation}
    d_N = \mathrm{Tr}(\mathbf{A}^T\mathbf{D}_N\mathbf{A}), \mathbf{D}_N = \sum_{i=1}^{c_s}\big(\frac{1}{|N_i|} \sum_{k\in{N_i}}\Delta \boldsymbol{x}_{N}^k{(\Delta \boldsymbol{x}_{N}^k)}^T\big)
\end{equation}
Similarly, 
\begin{equation}
    d_F = \mathrm{Tr}(\mathbf{A}^T\mathbf{F}_N\mathbf{A}), \mathbf{F}_N = \sum_{i=1}^{c_s}\big(\frac{1}{|F_i|} \sum_{k\in{F_i}}\Delta \boldsymbol{x}_{F}^k{(\Delta \boldsymbol{x}_{F}^k)}^T\big)
\end{equation}
where $F_i = \{k|k\ \mathrm{is\ not\ the\ spatial\ neighbor\ of}\ i\}$ and $\Delta \boldsymbol{x}_{F}^k = \bar{\boldsymbol{x}}_s^i - \bar{\boldsymbol{x}}_s^k$ for $k \in F_i $.\\
\indent
Having defined metrics $d_N$ and $d_F$, we can introduce an inequality to keep the projected CSI distance of spatial neighboring points close, while driving that of 
other points far away:
\begin{equation}
    d_N \leq \alpha \cdot d_F.
    \label{eq:ineq}
\end{equation}

\subsubsection{Algorithm}
By incorporating the objectives in 
the above subsections, 

we formulate the optimization problem as maximizing $s_b$ while minimizing $s_t$ and $s_w$, and a regularization term $||\mathbf{A}||_F$. Due to the arbitrary of the absolute value of $\mathbf{A}$, it makes sense that we can fix $\mathbf{A}^{T}\mathbf{Q}_s\mathbf{A}$, whose trace is $s_b$, as the identity matrix $\mathbf{I}$ and maximize the other values. The optimal function $f(\mathbf{A})$ can be represented as:
\begin{align}
    \min_{\mathbf{A}}\  &f(\mathbf{A}) =s_t+\mu s_w+\lambda||\mathbf{A}||_F^{2},\\
    \text{s.t. }&
    \mathbf{A}^{T}\mathbf{Q}_s\mathbf{A}=\mathbf{I},
    \label{Eq1}
\end{align}
where $\mu$, $\lambda$ are the parameters for weighting the intra-class distance and the Frobenius norm. We plug Equation \eqref{eq:s_t}, \eqref{eq:s_w} and \eqref{eq:s_b} in Equation \eqref{Eq1}, obtaining:
\begin{align}
    \min_{\mathbf{A}}\ & f(\mathbf{A})=\mathrm{Tr}(\mathbf{A}^{T}\mathbf{X}\mathbf{M}\mathbf{X}^{T}\mathbf{A}+\mu\mathbf{A}^{T}\mathbf{P}_{s}\mathbf{A}+\lambda\mathbf{A}^{T}\mathbf{A})\nonumber\\
    \text{s.t. }& \mathbf{A}^{T}\mathbf{Q}_s\mathbf{A}=\mathbf{I}.
    \label{Eq2}
\end{align}
The Lagrangian approach is used to find the optimality of Equation \eqref{Eq2}, which is:
\begin{equation}
    \min_{\mathbf{A}} 
    \mathrm{Tr}(\mathbf{A}^{T}(\mathbf{X}\mathbf{M}\mathbf{X}^{T}+\mu\mathbf{P}_s+\lambda\mathbf{I})\mathbf{A})-\mathrm{Tr}((\mathbf{A}^{T}\mathbf{Q}_s\mathbf{A}-\mathbf{I})\mathbf{Z}),
    \label{Eq4}
\end{equation}
where $\mathbf{Z} = \text{diag} (z_1, z_2, \dots,\ z_M)$ is the Lagrangian multiplier with $M$ as the rank of the matrix $\mathbf{A}$. To find the minimum value, the derivatives $\frac{\partial f(\mathbf{A})}{\partial \mathbf{A}}$ should equal to a zero matrix, that is,
\begin{equation}
    (\mathbf{X}\mathbf{M}\mathbf{X}^{T}+\mu\mathbf{P}_s+\lambda\mathbf{I})\mathbf{A} = \mathbf{Q}_s\mathbf{A}\mathbf{Z}.
    \label{Eq3}
\end{equation}

By multiplying $\mathbf{A}^{T}$ on the left side of both sides of the Equation \eqref{Eq3} and submitting $\mathbf{Z}$ into Equation \eqref{Eq4}, we simplify the above optimality condition as:
\begin{equation}
    \min_{\mathbf{A}} \mathrm{Tr}((\mathbf{A}^{T}\mathbf{Q}_s\mathbf{A})^{-1}\mathbf{A}^{T}(\mathbf{X}\mathbf{M}\mathbf{X}^{T}+\mu\mathbf{P}_s+\lambda\mathbf{I})\mathbf{A}),
    \label{Eq5}
\end{equation}
\begin{equation}
    \text{or }\min_{\mathbf{A}}\mathrm{Tr}(\mathbf{Z}),
\end{equation}
which can be transferred as:

\begin{equation}
    \max_{\mathbf{A}}
    \mathrm{Tr}((\mathbf{A}^{T}(\mathbf{X}\mathbf{M}\mathbf{X}^{T}+\mu\mathbf{P}_s+\lambda\mathbf{I})\mathbf{A})^{-1}\mathbf{A}^{T}\mathbf{Q}_s\mathbf{A}),
    \label{Eq6}
\end{equation}
\begin{equation}
    \text{or }\max_{\mathbf{A}}\mathrm{Tr}(\mathbf{Z}^{-1}).
    \label{Eq7}
\end{equation}

We adopt an approach similar to kernel Fisher discriminant (KFD) analysis 
\cite{914517} to find the optimal solution. 

According to Equation \eqref{Eq3} we can derive:
\begin{equation}
    \mathbf{A}^{T}(\mathbf{X}\mathbf{M}\mathbf{X}^{T}+\mu\mathbf{P}_s+\lambda\mathbf{I})\mathbf{A}=\mathbf{A}^{T}\mathbf{Q}_s\mathbf{A}\mathbf{Z},
\end{equation}
from which we can get:
\begin{equation}
    \mathbf{A}\mathbf{Z}^{-1} = (\mathbf{X}\mathbf{M}\mathbf{X}^{T}+\mu\mathbf{P}_s+\lambda\mathbf{I})^{-1}\mathbf{Q}_s\mathbf{A}.
\end{equation}
Note that the inverse of Lagrange multiplier is $\mathbf{Z}^{-1} = \text{diag} (z_1^{-1}, z_2^{-1}, \dots,\ z_M^{-1})$, a diagonal matrix. For the $i$th column vector $\boldsymbol{a}_i$ in $\mathbf{A}$, there has:
\begin{equation}
    z_i^{-1}\boldsymbol{a}_i= (\mathbf{X}\mathbf{M}\mathbf{X}^{T}+\mu\mathbf{P}_s+\lambda\mathbf{I})^{-1}\mathbf{Q}_s\boldsymbol{a}_i
\end{equation}
which suggests that each $z_i$ is one of the eigenvalues w.r.t $(\mathbf{X}\mathbf{M}\mathbf{X}^{T}+\mu\mathbf{P}_s+\lambda\mathbf{I})^{-1}\mathbf{Q}_s$ and $\mathbf{A}$ is a combination of independent column eigenvectors. Considering Equation \eqref{Eq7}, the larger the trace of $\mathbf{Z}^{-1}$ is, i.e., the larger of the sum of selected eigenvalues is, the larger the optimal function $f(\mathbf{A})$ is. Therefore, 
the transform matrix $\mathbf{A}$ is made of $M$ largest independent eigenvalues of $(\mathbf{X}\mathbf{M}\mathbf{X}^{T}+\mu\mathbf{P}_s+\lambda\mathbf{I})^{-1}\mathbf{Q}_s$.

We further introduce the additional inequality Equation \eqref{eq:ineq} to refine 
the selection of the eigenvalues of $\mathbf{A}$. Then, the solution of $\mathbf{A}$ is the matrix with the largest corresponding eigenvalues that satisfies the inequality constraint.\\

%% file: 4_4_matching.tex
\subsection{Matching Rule}
\label{sec: localization}
We propose the edge enhanced $k$-nearest neighbors algorithm (EEKNN) as the matching rule for indoor localization. It specially tackles the decreasing accuracy of 
fingerprinting based localization on the edge or at the corner,  where there are fewer training points as neighbors for the testing points.
\subsubsection{Motivation}
EEKNN is derived from weighted $k$-nearest neighbors (WKNN) and motivated by the following two observations.\\
\begin{figure}[t]
	\centering
	\setlength\abovecaptionskip{-1pt}
	\setlength\belowcaptionskip{-1pt}
	\begin{minipage}[t]{1\linewidth}
		\centering
		\subfloat[CSI]
		{\includegraphics[width=0.45\textwidth]{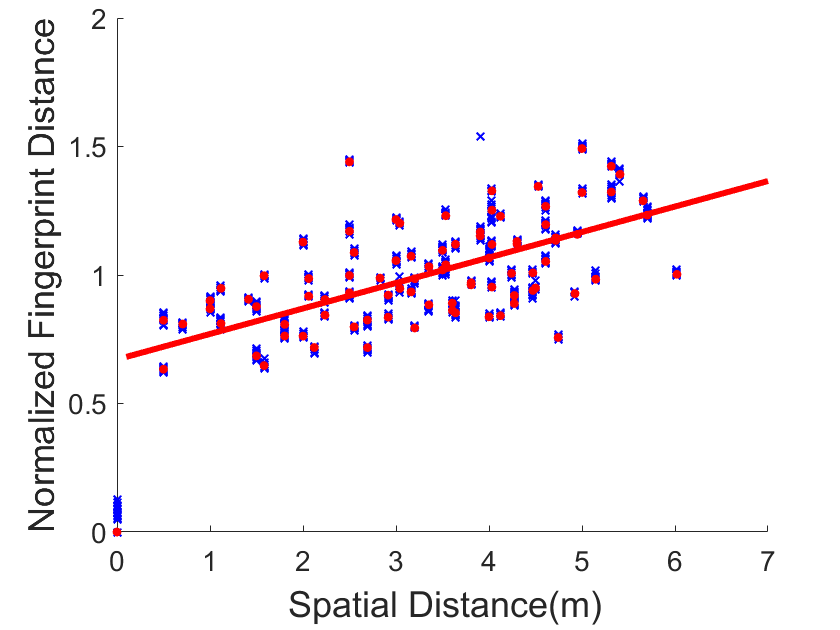}\label{fig: chick CSI 1}}
		\hspace{0.05cm}
		\subfloat[RSS]
		{\includegraphics[width=0.45\textwidth]{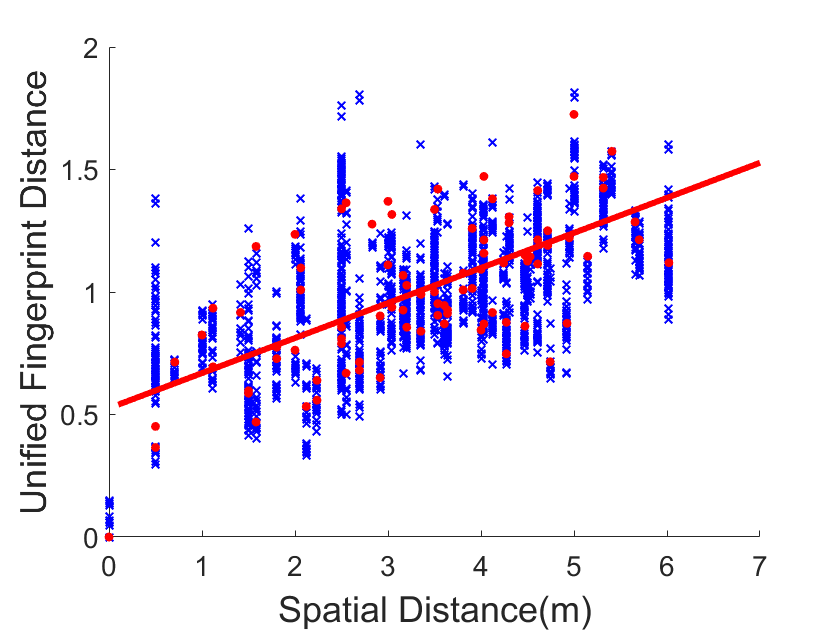}\label{fig: chick RSSI 1}}
		\hspace{0.05cm}
		\subfloat[CSI]
		{\includegraphics[width=0.45\textwidth]{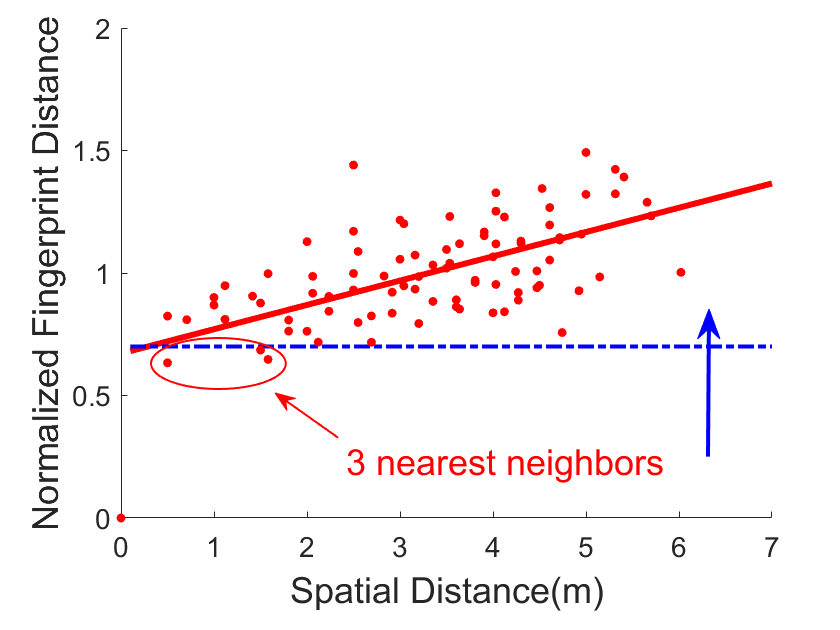}\label{fig: chick CSI 2}}
		\hspace{0.05cm}
		\subfloat[RSS]
		{\includegraphics[width=0.45\textwidth]{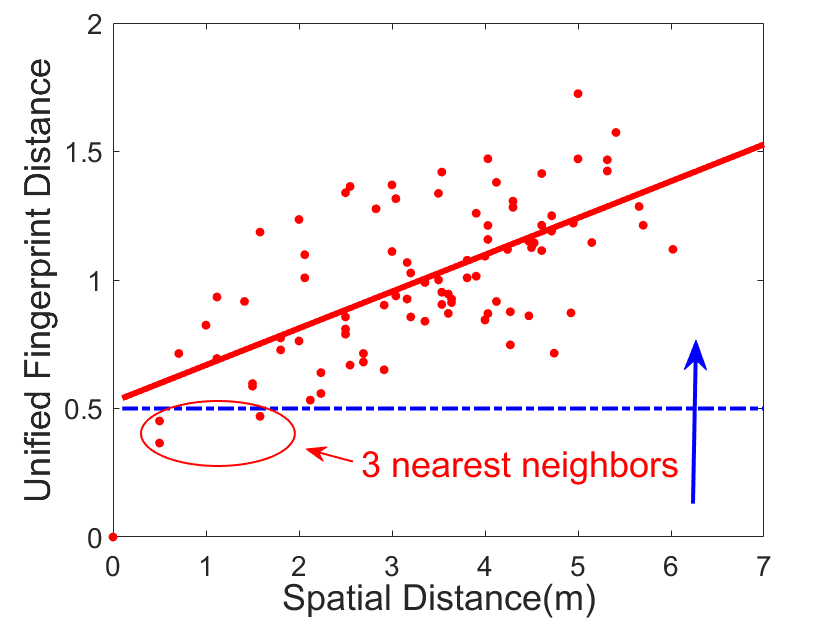}\label{fig: chick RSSI 2}}
		\hspace{0.05cm}
		\caption{Euclidean distances of CSI/RSS vectors versus spatial distances of all training points with respect to a reference position.
		The blue crosses represent the multiple samples in each training points and the red circles represent the average of multiple samples for each training point. 
		Red lines are drawn as the trend lines of red circles using least squares method.
		Both CSI and RSS are mean-normalized.}
		\label{fig: chick}
	\end{minipage}
\end{figure}

\textbf{Profound Perspective of CSI}. We hereby provide a profound perspective of the role that CSI plays in localization. 
In the Fig.\ref{fig: chick}, we plot the relationship between the spatial distances and fingerprint distances of all training points with respect to a certain location. Yellow circles are generated by the mean value of all samples in one position, and blue ones are generated by samples. The red line is the linear regression using least squares.\\ 
\indent
Compared with RSS, CSI is much more stable. Blue circles of RSS are scattered in Fig.\ref{fig: chick RSSI 1} while those of CSI cluster around the average (yellow circles) in Fig.\ref{fig: chick CSI 1}. Such stability is beneficial to localization.\\
\indent
On the other hand, the fingerprint differentials across spatial distance of CSI is comparatively low, which serves as a deficiency of CSI. The slope of the red line of CSI is not as steep as that of RSS. For example, Fig.\ref{fig: chick CSI 2} and Fig.\ref{fig: chick RSSI 2} demonstrate the process of seeking the three nearest neighbors: the dash-dotted blue line moves upwards until three neighbors are found. Due to the gentle slope, the spatial distances of CSI neighbors tend to be much farther than those of RSS, thus reducing the localization accuracy. That is to say, the number of neighbors $k$ in WKNN has more critical influence on CSI-based localization, especially for edge points where there are fewer spatial nearest neighbors.

\begin{figure}[t]
    \centering
    \includegraphics[width=0.88\columnwidth]{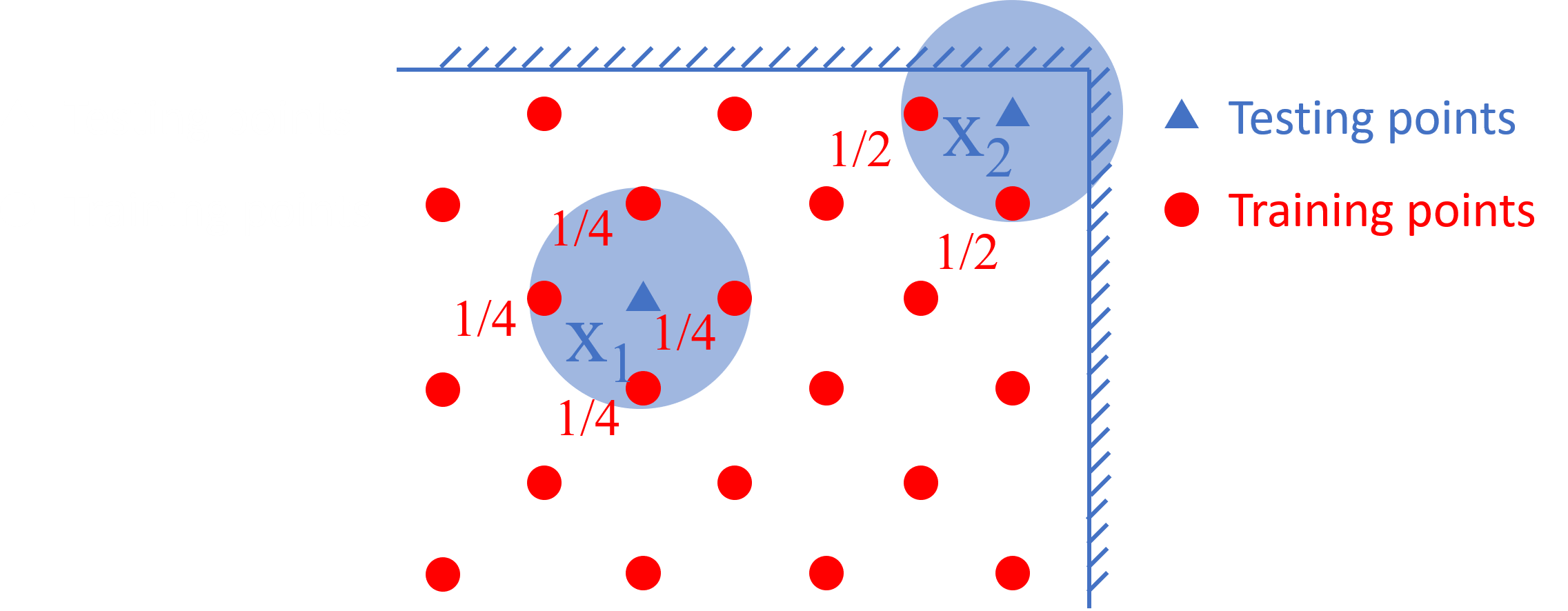}
    \caption{An example of nearest neighbors: The central point has 4 physical nearest neighbors, while the corner point has only two.}
    \label{fig: toy example for localization}
\end{figure}

\textbf{Corner points suffer larger errors}. When implementing WKNN in our system, we observe that the predicted positions of corner points are usually farther apart from the corner than they actually are. The reason is that all the neighbors of a corner point are on its single side, thus pulling the prediction result in one direction away from the corner. Even worse, unlike central points, a corner point has only two spatially nearest neighbors (Fig.\ref{fig: toy example for localization}). When $k$ is greater than two, WKNN may pick up one far-away training point as a neighbour, thus incurring big errors.\\

\textbf{Algorithm}. Edge enhanced $k$-nearest neighbors (EEKNN) is a method purposed based on these two observations. It improves the accuracy of edge and corner points while maintaining the accuracy of central points simultaneously. The main idea is to automatically adjust the number of neighbors $k$ and the weights of different neighbors on the basis of their spatial locations. Recall that edge and corner points are prone to having fewer nearest neighbors. Hence, we decrease the number of neighbors once corner (or edge) training points are selected. Moreover, these points are given higher weights so as to pull the predictions back to the edge.\\
\indent
The algorithm is illustrated as follows. We refer to the inverse of the number of nearest neighbors $N^{(i)}_{neighbor}$ that training point $i$ has as neighbor portion $\kappa_i$: \\
\begin{equation}
    \begin{aligned}
        \kappa_i=\frac{1}{N^{(i)}_{neighbor}}.
    \end{aligned}
\end{equation}
\indent
For each testing point, we find several neighbors i with $\kappa_i$ sum up to $k'$. Here, $k'$ equals 1 in the default settings. Then we weight these training points not only by the fingerprint distances, but $\kappa_i$ as well. The larger $\kappa_i$ is, the higher proportion training point $i$ accounts for among all neighbors.\\
\begin{equation}
\begin{aligned}
& \boldsymbol{\hat{p}} = \sum_{i}w_i\ \boldsymbol{p}_i\\
& w_i=\frac{1/(\varepsilon_i\times \kappa_i)}{\sum_j{1/(\varepsilon_j\times \kappa_j)}}\\
\end{aligned}
\end{equation}
where $\varepsilon_i$ is the Euclidean distance as it is in WKNN, $\boldsymbol{\hat{p}}$ is the estimated location, and $\boldsymbol{p}_i$ is the position of the $i$th neighbor.\\
\indent
Under such definition, edge or corner training points are assigned a larger $\kappa_i$ than central ones. As for the testing points, those on the edges are more likely to pick up neighbors with a larger $\kappa_i$. For example in Fig. \ref{fig: toy example for localization}, central points with four nearest training point neighbors tend to have $\kappa_i = 1/4$ for four neighbors while the corner points probably pick up only two nearest neighbors, whose $\kappa_i$'s equal $1/2$. Hence, the numbers of neighbors for central points remain and that for edge and corner points goes down.\\

%% file: 5_setup.tex
\section{Experimental Setup}
\label{sec: setup}
In this section, we describe the software and hardware configurations for 
CSI fingerprinting, and the performance metrics for indoor localization. 

\subsection{Devices}
\begin{figure}[t]
    \centering
    \includegraphics[width=0.5\columnwidth]{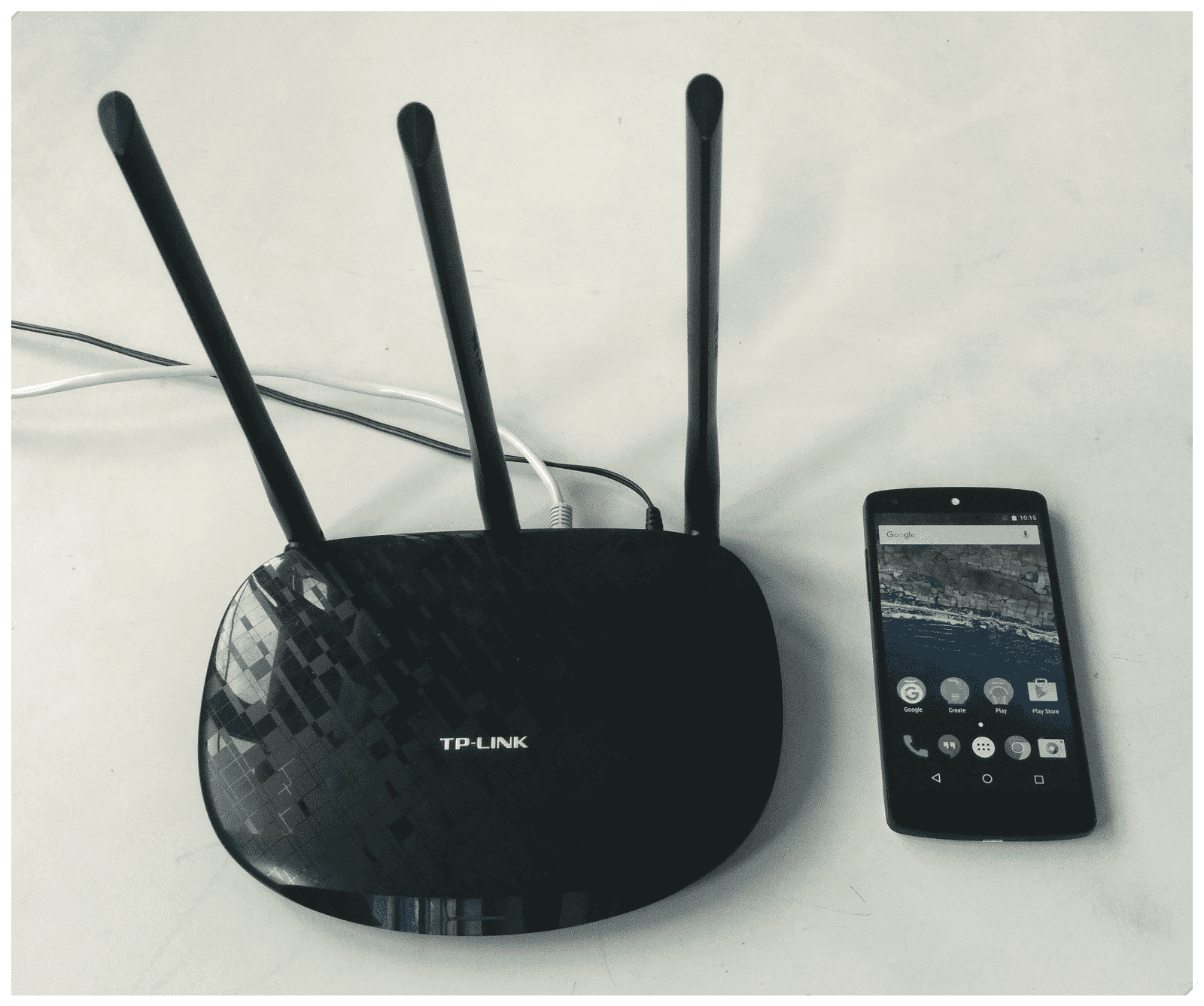}
    \caption{The Nexus 5 smart phone and the router.}
    \label{fig: nexus5}
\end{figure}

We implement CRISLoc using TPlink TL-WR885N as routers and Nexus 5 as mobile devices. The whole system works at 2.4 GHz with the bandwidth of 20MHz. Moreover, we select a relative empty channel to avoid the interference from adjacent traffic.\\
\indent
\textbf{RSS}. We develope an Android application to extract RSS from multiple APs simultaneously with an interface provided by Android Studio. The extracted RSS is represented by decibels, ranging from -100 to 0 dBm. For those routers whose signals are too weak to be detected by our android APP, we set RSS as a minimum value -100 dBm by default.\\
\indent
\textbf{CSI}. Nexus 5 with Nexmon\cite{nexmon:project} installed overhears frames transmitted by the router and extracts the CSI from the router to the Nexus 5, approximately up to 100 frames per second. The rate decreases as the distance between Nexus 5 and router increases. As for the cases that the routers and the smartphones are so far apart that the frames cannot be caught, we set CSI as 0 in every subcarrier.\\

\vspace{-0.45cm}
\subsection{Scenarios}
\begin{figure}[t]
	\centering
	\setlength\abovecaptionskip{-1pt}
	\setlength\belowcaptionskip{-1pt}
	\begin{minipage}[t]{1\linewidth}
		\centering
		\subfloat[Research laboratory]
		{\includegraphics[width=0.7\textwidth]{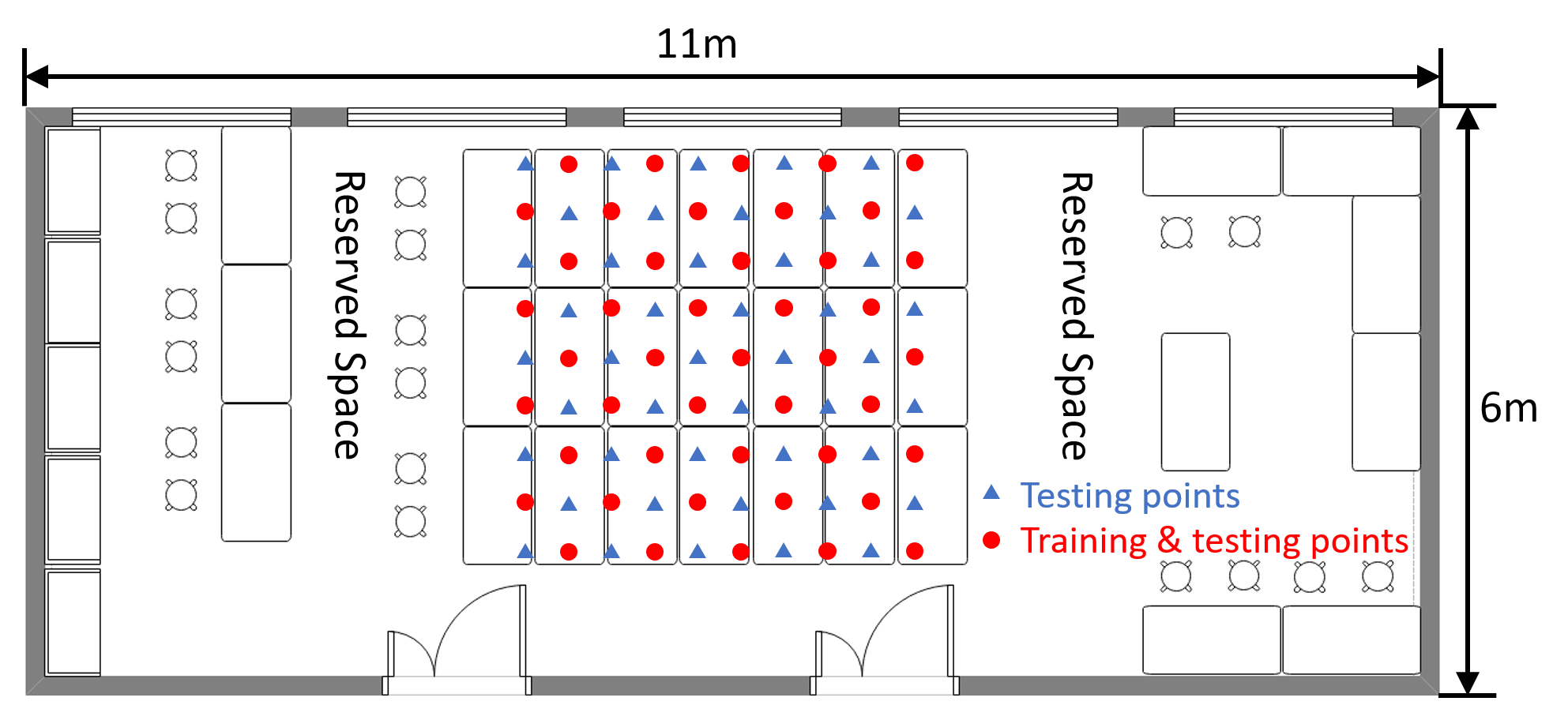}\label{fig: lab}}
		\hspace{0.05cm}
		\subfloat[Academic building]
		{\includegraphics[width=0.9\textwidth]{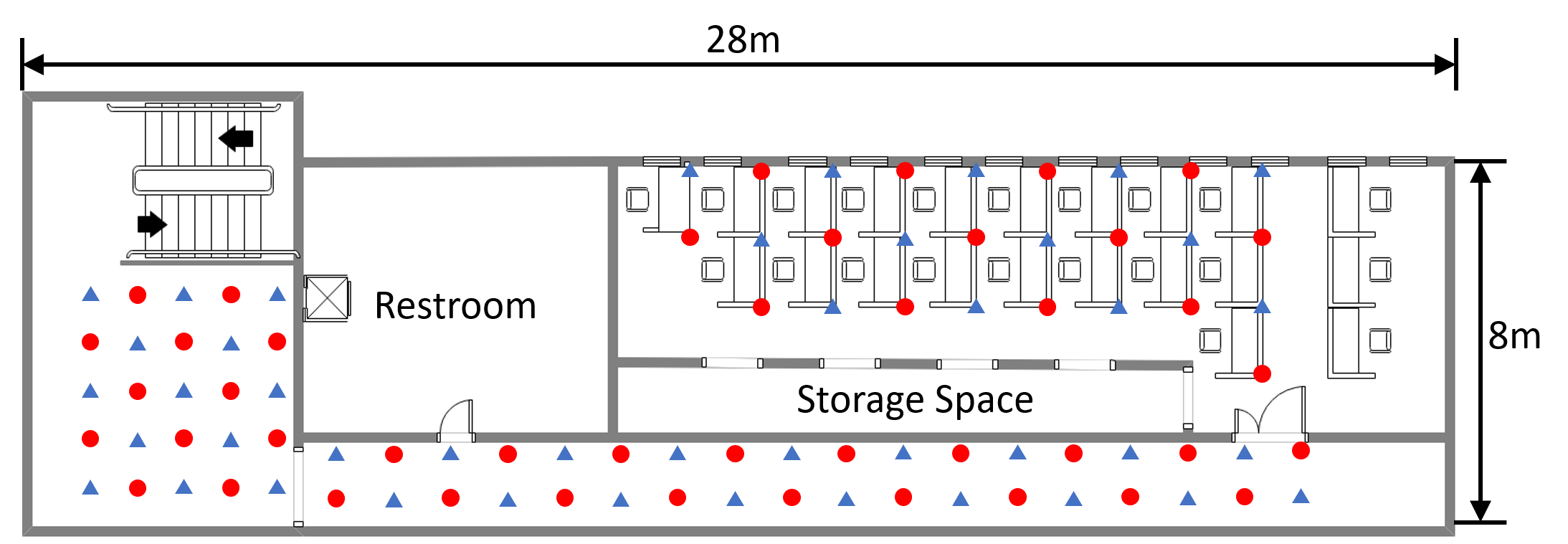}\label{fig: corridor}}
		\hspace{0.05cm}
		\caption{Floor plan{\color{blue}, where triangle points represent testing points and circle points represent training and testing points}.}
		\label{fig: floorplan}
	\end{minipage}
\end{figure}

We implement CRISLoc in two different typical indoor localization scenarios:\\
\indent
\textbf{Research laboratory}. First, we set up a testbed in the center of a $6m\times 15m$ research laboratory on the desks (Fig.\ref{fig: lab}). There are few multipath reflections and little disturbance. RSS and CSI are collected at 90 positions with $0.5m$ spacing. Nine APs are placed inside the room. \\
\indent
\textbf{Academic Building}. Then, we conduct the experiment on the third floor of an academic building (Fig. \ref{fig: corridor}). The test area is much more complex with many obstacles around. People walk around when fingerprints are collected, thus bringing disturbance to the data. The area covers an office, a corridor, and a lobby, and it is divided into grids with the edge width of $1.2m$ on average at the height ranging from 0.8m to 1.5m. Ten APs in all are deployed: five in the corridor and five in the office.\\

\indent
Points marked with circles in Fig.\ref{fig: floorplan} are used as both training points in site survey and testing points, where there are eight reference points that distribute evenly. The other half of the points, those marked with triangles, are used as testing points only. The training points and testing points are distributed alternatively.\\

\subsection{Evaluation metrics}
Different from traditional multi-class classifications in machine learning that only one class will be selected as the results, altered AP detection does not know whether or what number of APs are altered: there are cases where either none alters or multiple APs alter. Hereby, we refer and extend the concepts Precision at K in the recommendation systems as our evaluation metrics\cite{powers2011evaluation}.\\

\indent
\textbf{Relevant and Recommended.}
In our implementation, we define Relevant items  as the altered APs, noted as $S_{rel}$. Recommended items represent those alarmed APs, marked as $S_{rec}$. Different from the traditional recommendation systems, the number of Recommended items is not given by the user but by our algorithm automatically. 

\indent
\textbf{Precision, Recall and F1-score at K.}
{\color{blue}We refer the original concept of Precision at K in the recommendation systems, which describes the proportion of the altered APs among alarmed APs.}
\begin{equation}
    Precision@k = \frac{|S_{rel} \cap S_{rec}|}{|S_{rec}|}.
\end{equation}

We are  interested in the Recall at K, which evaluates the proportion of the detected APs among all altered APs,
\begin{equation}
    Recall@k = \frac{|S_{rel} \cap S_{rec}|}{|S_{rel}|}.
\end{equation}

Still, F1-score at K is the harmonic mean of precision and recall, calculated as
\begin{equation}
    \frac{precision \cdot recall}{(precision + recall) / 2}.
\end{equation}
\\

%% file: 6_1_matching.tex
\section{Experimental Evaluation}
\label{sec: results}
In this section, we begin with a set of micro-benchmark experiments to 
validate the effectiveness of CSI calibration and EEKNN. We next 
evaluate the performance of altered AP detection and fingerprint reconstruction separately, and CRISLoc as a unity.

\subsection{Effectiveness of CSI Calibration and EEKNN}

\begin{figure}[t]
	\centering
	\setlength\abovecaptionskip{-1pt}
	\setlength\belowcaptionskip{-1pt}
	\begin{minipage}[t]{1\linewidth}
		\centering
		\subfloat[Research laboratory]
		{\includegraphics[width=0.48\textwidth]{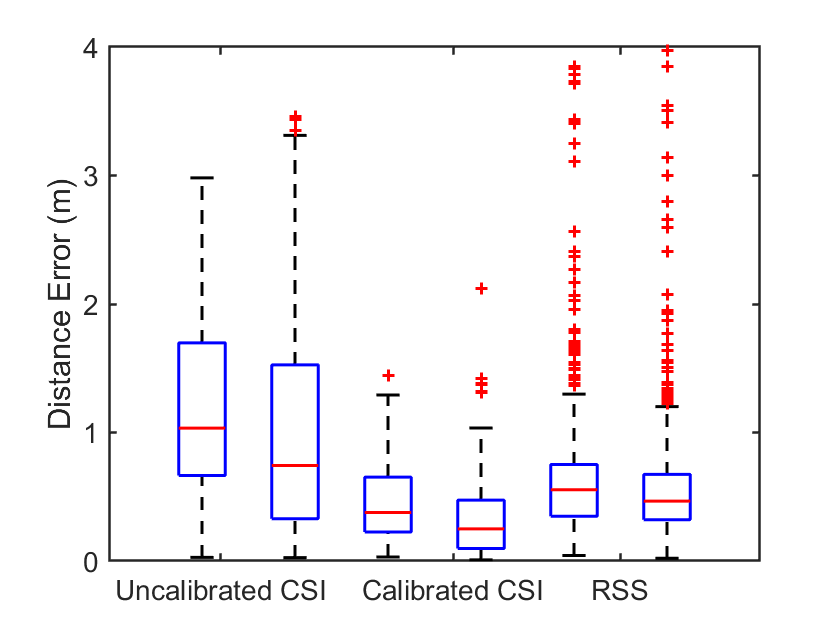}\label{fig: box 625}}
		\hspace{0.05cm}
		\subfloat[Academic building]
		{\includegraphics[width=0.48\textwidth]{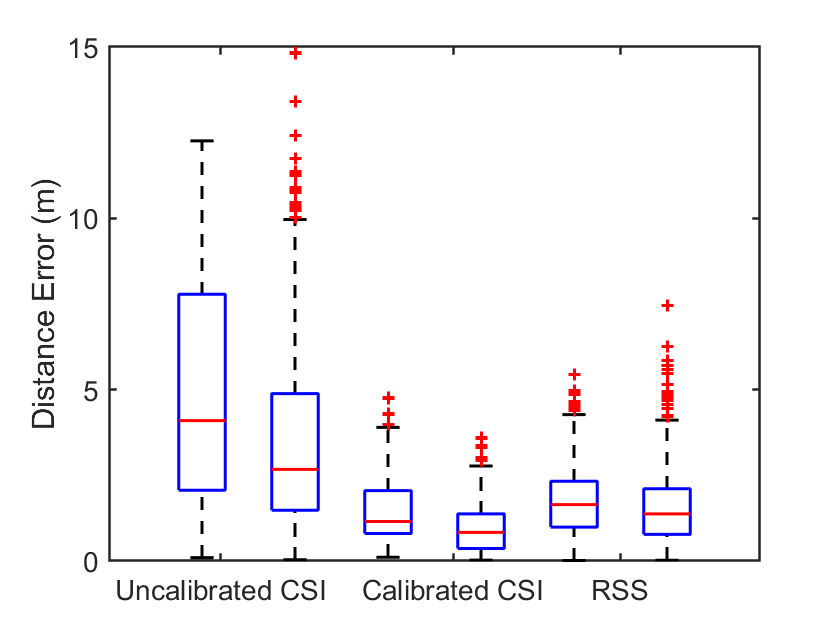}\label{fig: box 319}}
		\caption{Localization results of uncalibrated CSI, calibrated CSI, and RSS: for every pair in the figure, the left uses WKNN while the right uses EEKNN.}
		\label{fig: box}
	\end{minipage}
\end{figure}

\indent
In this set of experiments shown in Fig.\ref{fig: box}, we evaluate the accuracy of localization using different forms of fingerprints, namely, uncalibrated CSI, calibrated CSI, and RSS. Uncalibrated CSI yields poor results because it does not follow the rule that closer points share similar fingerprints due to the automatic gain control. Calibrated CSI performs better than RSS in two aspects. First, the overall error of calibrated CSI is lower than that of RSS: the mean distance error of CSI based localization is 36.1\% lower than that of RSS in the research laboratory, and 35.6\% lower in the academic building. {\color{blue}Second, fewer outliers over the boxes of CSI-based localization appear, due to the high stability of calibrated CSI with fewer abnormal measurements.} 
Fig.\ref{fig: box} demonstrates the effectiveness of EEKNN as well. By using an adaptive neighbor portion $\kappa_i$, EEKNN avoids picking up a physically far-away neighbour and greatly improves localization accuracy, especially for corner points and edge points. In the academic building with more corners and edges, the mean distance error of calibrated CSI is reduced by 34.1\% while the error is reduced by 21.3\% in the research laboratory. Besides, since CSI is more likely to pick up a far-away neighbor as $k$ goes up, EEKNN benefits calibrated CSI more than RSS. The mean error of RSS decreases by 11.6\% in the building and 11.2\% in the lab.

%% file: 6_2_detection.tex
\subsection{Altered AP Detection}
\begin{figure}[t]
	\centering
	\setlength\abovecaptionskip{-1pt}
	\setlength\belowcaptionskip{-1pt}
	\begin{minipage}[t]{1\linewidth}
		\centering
		\subfloat[Research laboratory]
		{\includegraphics[width=0.7\textwidth]{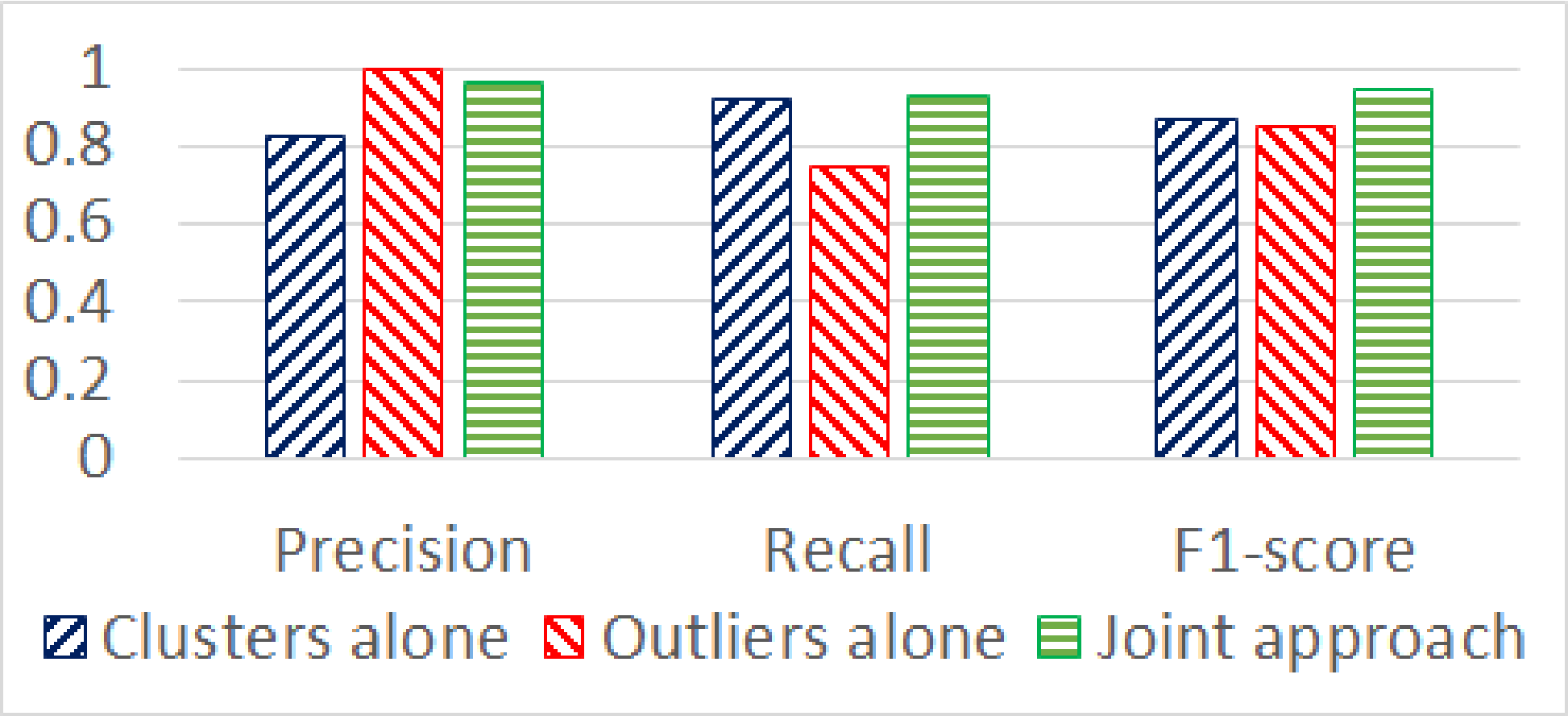}\label{fig: detection compare floor6}}
		\hspace{0.1cm}
		\subfloat[Academic building]
		{\includegraphics[width=0.7\textwidth]{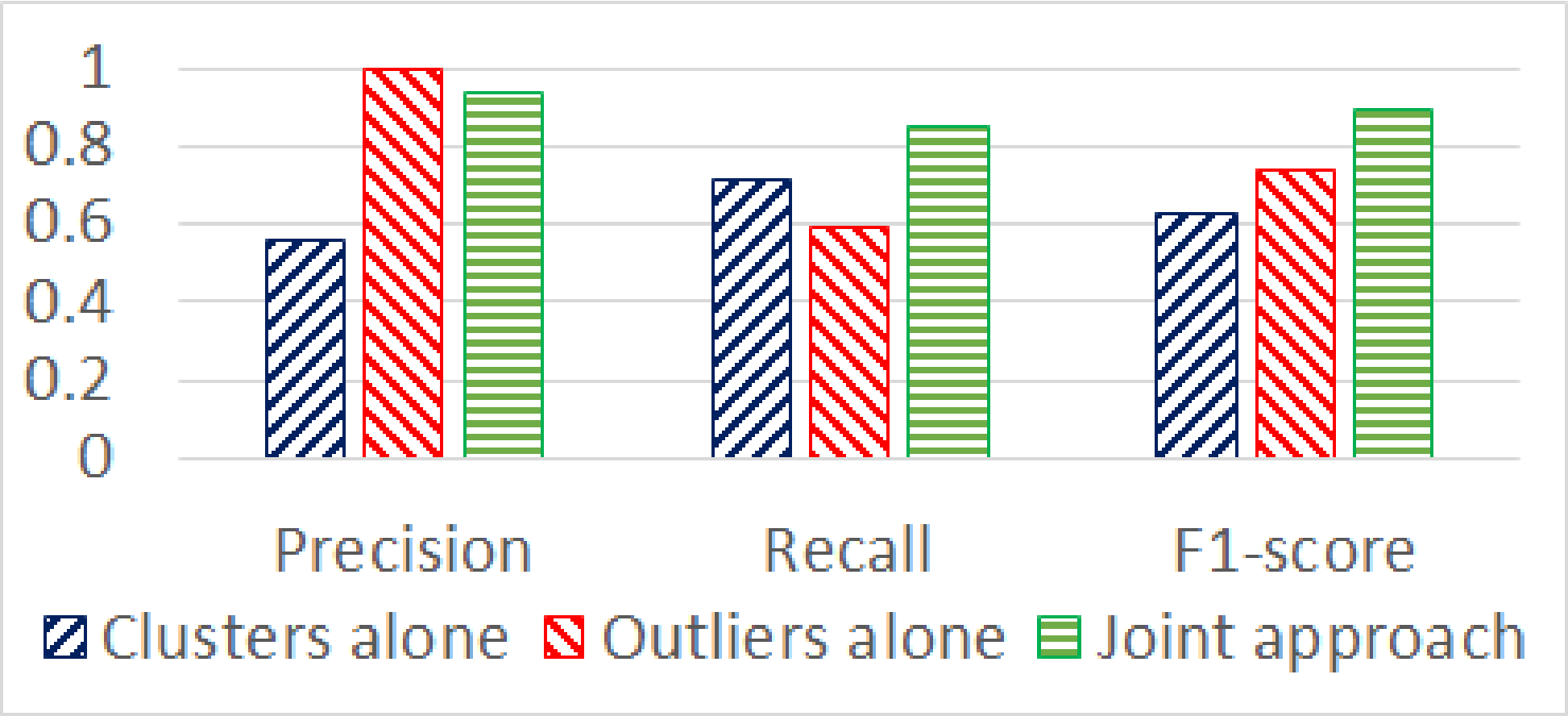}\label{fig: detection compare floor3}}
		\hspace{0.05cm}
		\caption{Detection without RPs by using different detection approaches}
		\label{fig: detection compare}
	\end{minipage}
\end{figure}

We first compare the improvement of joint approach against clustering or outlier alone shown in Fig.\ref{fig: detection compare}. As what we expect, clustering alone \cite{Cluster} suffers lower precision and outlier alone suffers lower recall, which is more intense in the academic building. The joint approach avoids the disadvantages of them, achieving the best F1-score. Fig.\ref{fig: detection compare} also shows that the performance in research laboratory is a little better than that in the academic building in general. \\

\begin{figure}[t]
	\centering
	\setlength\abovecaptionskip{-1pt}
	\setlength\belowcaptionskip{-1pt}
	\begin{minipage}[t]{1\linewidth}
		\centering
		\subfloat[Research laboratory]
		{\includegraphics[width=0.7\textwidth]{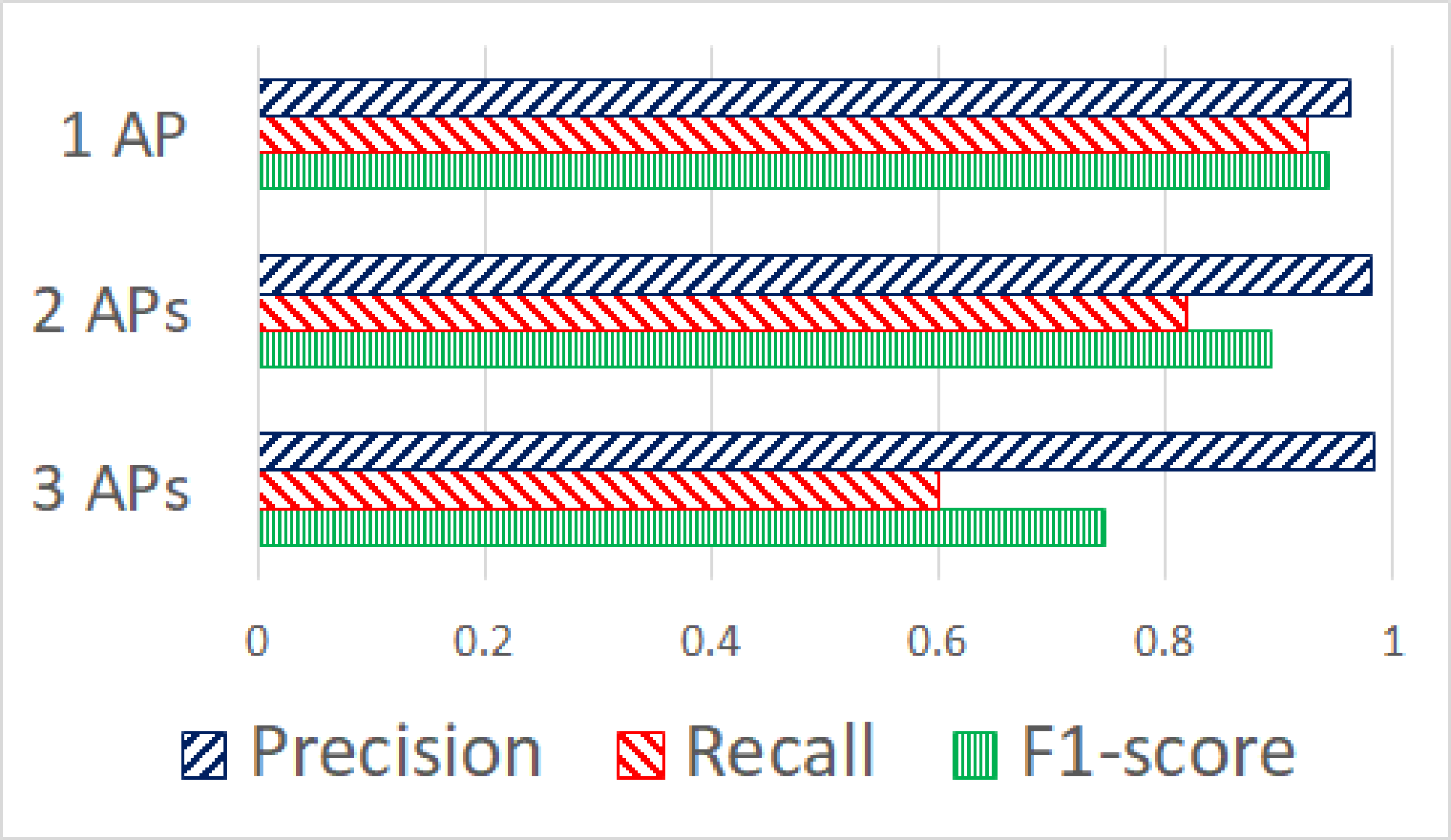}\label{fig: detection num floor6}}
		\hspace{0.1cm}
		\subfloat[Academic building]
		{\includegraphics[width=0.7\textwidth]{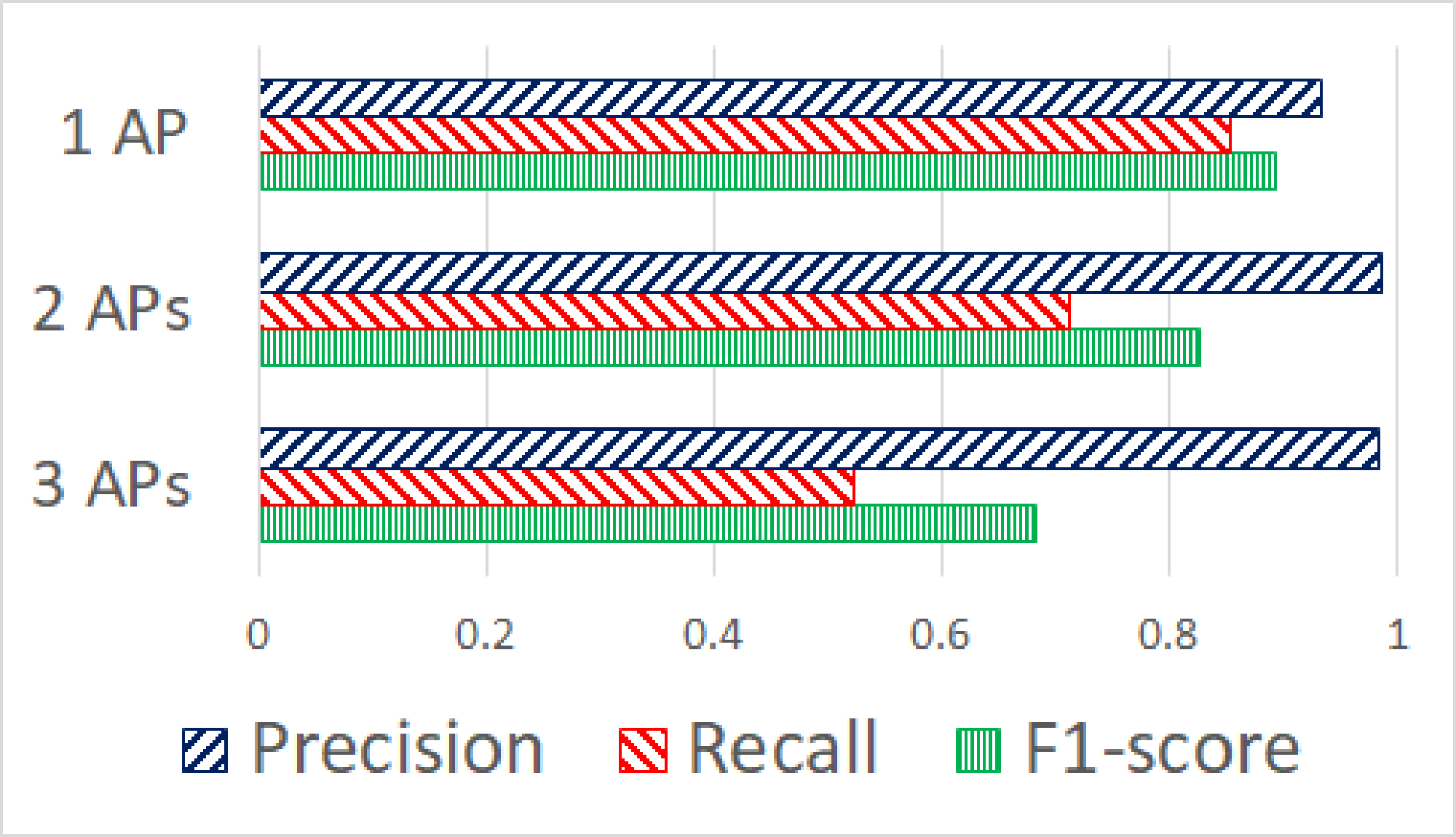}\label{fig: detection num floor3}}
		\hspace{0.05cm}
		\caption{Impact of the number of altered APs, detected without RPs by the joint approach.}
		\label{fig: detection num}
	\end{minipage}
\end{figure}

\indent
\indent
Then, we illustrate the impact of the number of altered APs in Fig.\ref{fig: detection num}. We claim that it is less likely in reality that multiple APs alter simultaneously, so we only examine the cases that one, two or three APs alter. It is inevitable that the precision, recall and F1-score decrease slighty as the number of altered APs increases, the F1-scores of which are still above 68\%.\\

\indent

%% file: 6_3_transfer.tex
\subsection{Fingerprint Reconstruction}
In this section, we evaluate the performance of fingerprint reconstruction separately, i.e., assuming that the prediction of altered AP is correct. We demonstrate its performance in two aspects: the error reduction by reconstruction and the impact of the number of RPs.
\subsubsection{Error Reduction by Fingerprint Reconstruction}

\label{sec: tf_res}

\begin{figure}[t]
	\centering
	\setlength\abovecaptionskip{-1pt}
	\setlength\belowcaptionskip{-1pt}
	\begin{minipage}[t]{1\linewidth}
		\centering
		
		\subfloat[Research laboratory $\times$ 1 AP]
		{\includegraphics[width=0.5\textwidth]{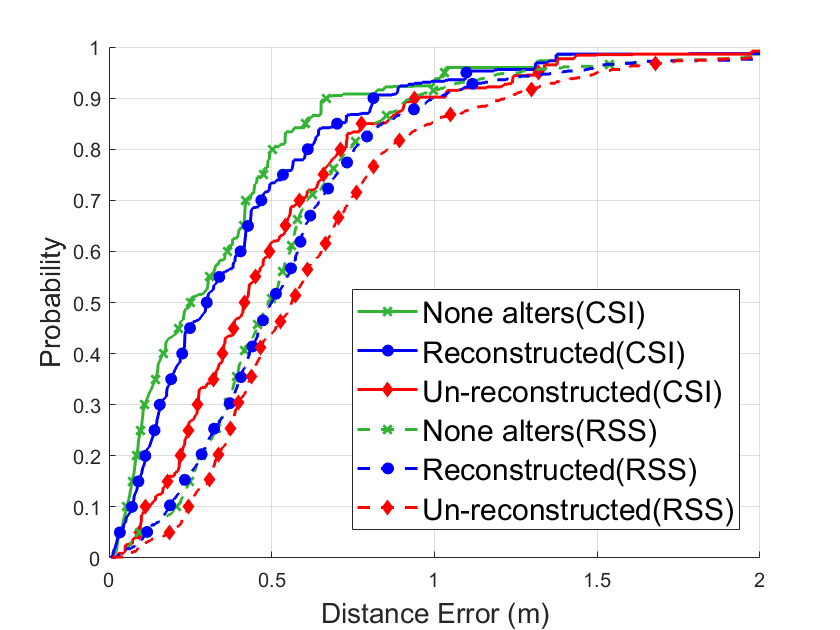}\label{fig: transfer 625 1}}
		\subfloat[Academic building $\times$ 1 AP]
		{\includegraphics[width=0.5\textwidth]{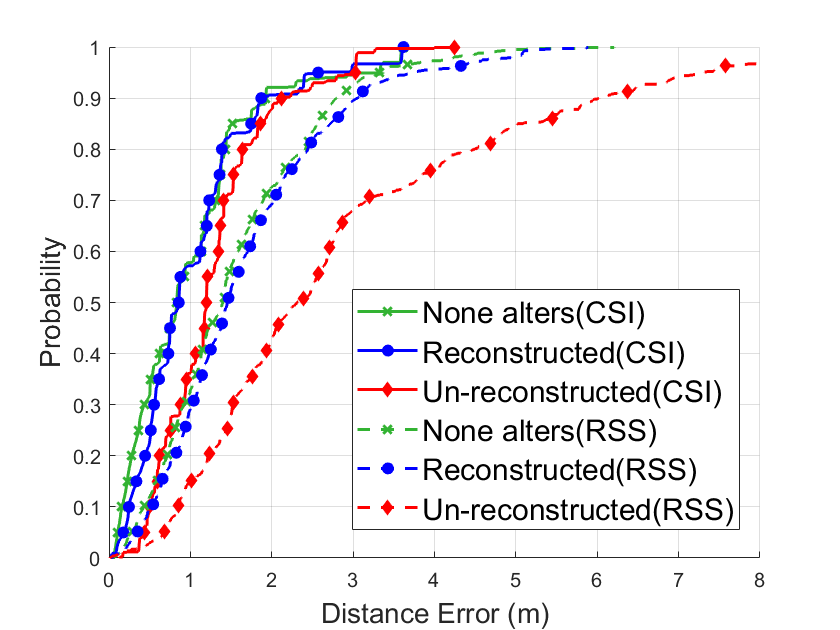}\label{fig: transfer 319 1}}
		
		\subfloat[Research laboratory $\times$ 2 APs]
		{\includegraphics[width=0.5\textwidth]{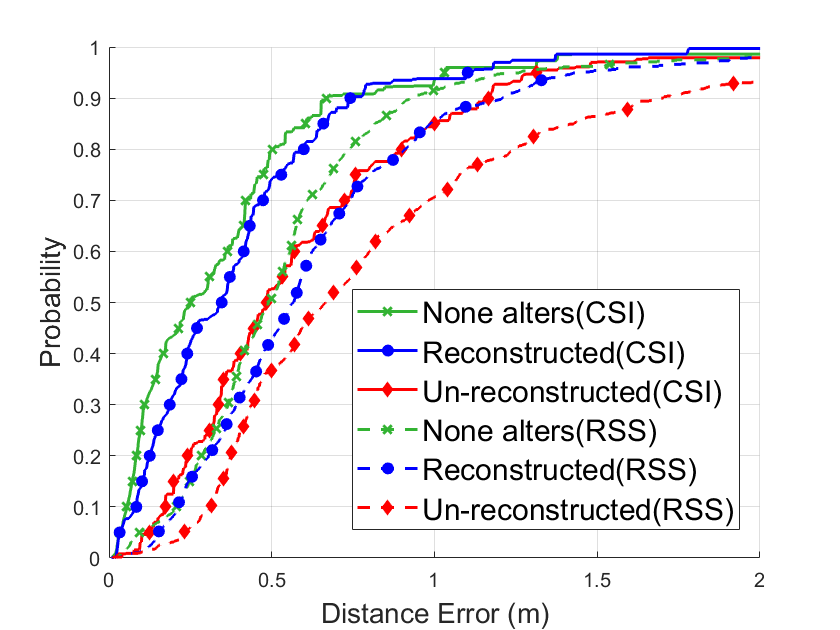}\label{fig: transfer 625 2}}
		\subfloat[Academic building $\times$ 2 APs]
		{\includegraphics[width=0.5\textwidth]{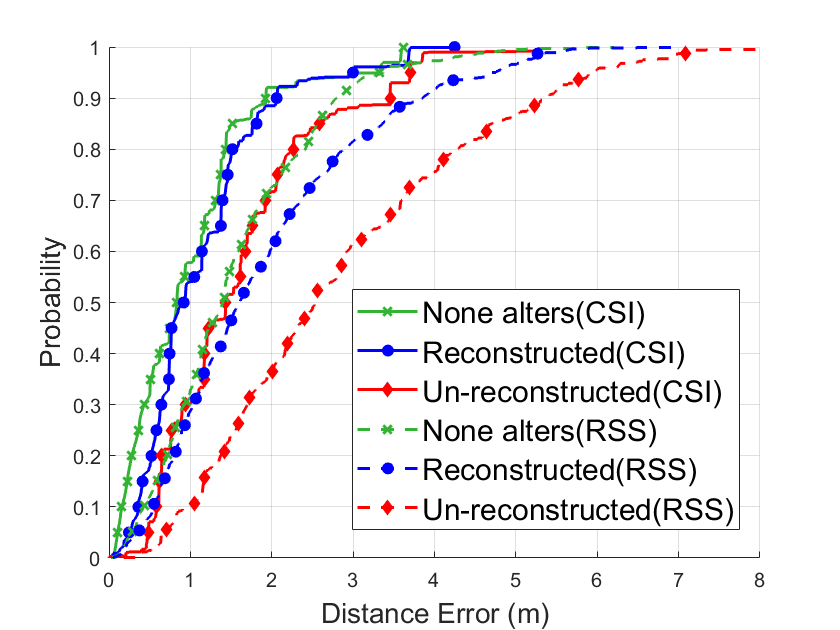}\label{fig: transfer 319 2}}
		
		\subfloat[Research laboratory $\times$ 3 APs]
		{\includegraphics[width=0.5\textwidth]{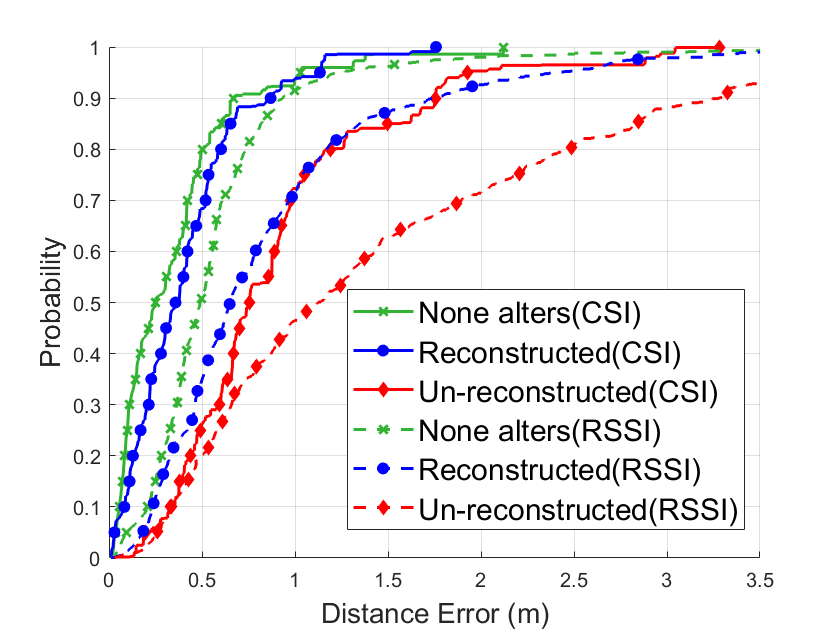}\label{fig: transfer 625 3}}
		\subfloat[Academic building $\times$ 3 APs]
		{\includegraphics[width=0.5\textwidth]{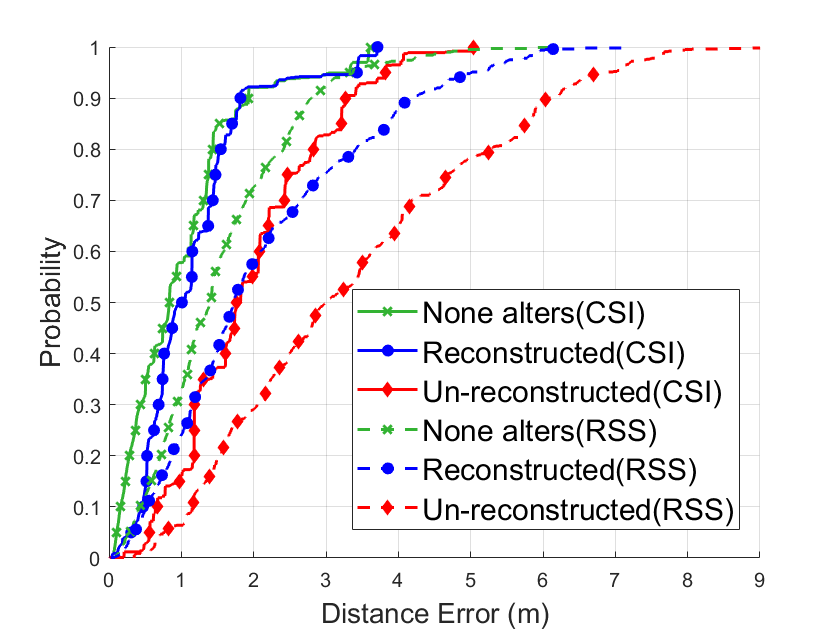}\label{fig: transfer 319 3}}
		
		\caption{The CDF of localization errors when 1, 2 and 3 AP(s) are altered at research laboratory and academic building. In each case, three solid lines represent the CDF of CSI in original, un-reconstructed and reconstructed situations respectively, and three dash lines represent that of RSSI.}
		\label{fig: transfer results}
	\end{minipage}
\end{figure}

\noindent
We hereby assess the performance of fingerprint reconstruction with eight RPs. There are three advantages in our reconstruction shown in Fig.\ref{fig: transfer results}.\\
\indent
First, CRISLoc eases the localization errors when a fraction of APs are altered. Particularly, when single AP is altered, which occurs the most frequently, the mean error is reduced from 0.46m (using out-of-date database) to 0.30m in the research lab, 
only a little higher than that in the initial situation 0.28m where no AP alters.\\
\indent
Second, the fingerprint reconstruction manages to reduce the dramatically increasing errors as the number of APs increases.\\

\indent
Third, CSI based localiztion outperforms RSS based localization no matter whether the fingerprint is newly collected, out-dated, or reconstructed. The dashed lines in Fig.\ref{fig: transfer results} demonstrated the efficiency of a similar RSS based fingerprinting localization system. Such RSS based localization reconstructs its fingerprints by LAAFU\cite{Cluster}, which requires a stringent condition that the fingerprints follow the path loss model. LAFFU's method cannot be applied to CSI, since the CSI amplitude of each subcarrier itself does not follow the path loss model as RSS does. Even worse, LAFFU even fails when the environment is complex and RSS does not follow this model. Transfer learning, as a result, conducts a better result.\\

To understand how the fingerprint reconstruction is influenced by the number of RPs, we measure the accuracy of fingerprints reconstructed with one to ten evenly-distributed RPs in the research lab in Fig.\ref{fig: RPnum}. Note that none represents the situation without reconstruction and inf represents the initial situation before AP alters. The reduction of error is quite obvious when the number of RPs increases from zero to four (no more than 9\% of the number of training points) in the research laboratory. This proportion varies with the complexity of the environment. 

\subsubsection{Impacts of the Number of RPs}
\begin{figure}[t]
    \centering
    \includegraphics[width=0.75\columnwidth]{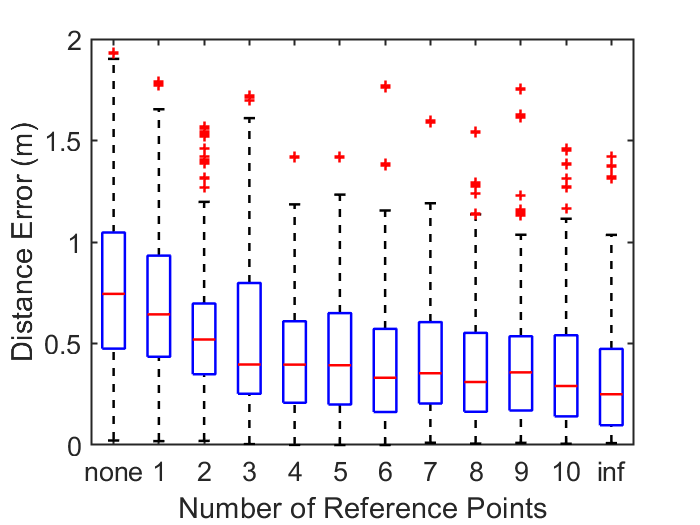}
    \caption{Impacts of the number of RPs on fingerprint reconstruction.}
    \label{fig: RPnum}
\end{figure}

To understand how the fingerprint reconstruction is influenced by the number of RPs, we measure the accuracy of fingerprints reconstructed with one to ten evenly-distributed RPs in the research lab in Fig.\ref{fig: RPnum}. Note that none represents the situation without reconstruction and inf represents the initial situation before AP alters. The reduction of error is quite obvious when the number of RPs increases from zero to four (no more than 9\% of the number of training points) in the research laboratory. This proportion varies with the complexity of the environment. 

%% file: 6_4_overall.tex
\subsection{Overall Performance}
\label{sec: overall}

\begin{figure}[t]
	\centering
	\setlength\abovecaptionskip{-1pt}
	\setlength\belowcaptionskip{-1pt}
	\begin{minipage}[t]{1\linewidth}
		\subfloat[Research laboratory.]
		{\includegraphics[width=0.5\textwidth]{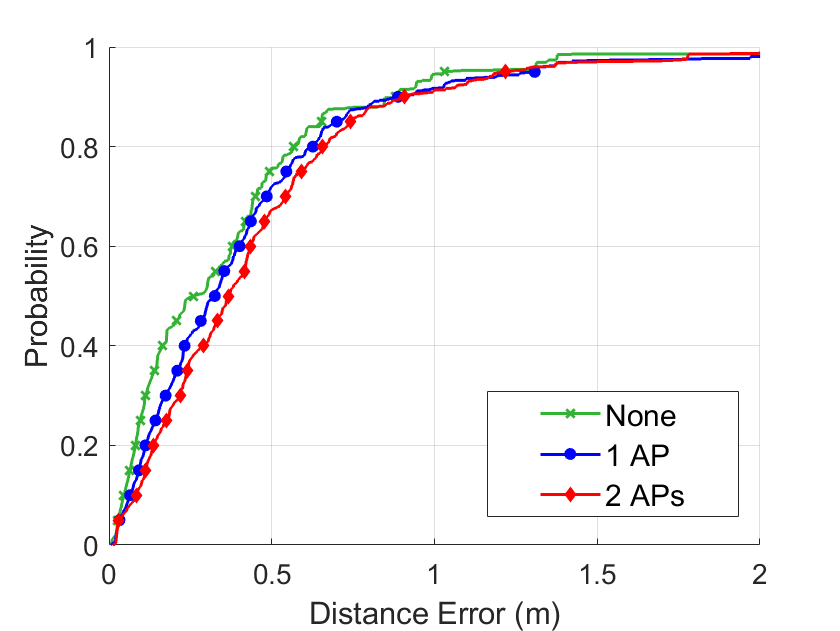}\label{fig: overall625}}
		\subfloat[Academic building.]
		{\includegraphics[width=0.5\textwidth]{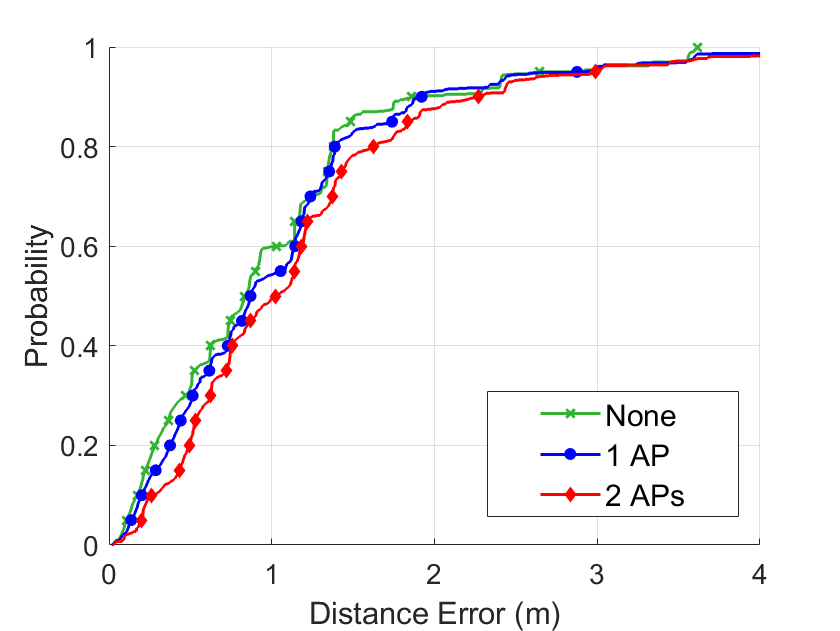}\label{fig: overall319}}
		\caption{Overall performance.}
		\label{fig: overallres}
	\end{minipage}
\end{figure}

The overall performance in shown in Fig.\ref{fig: overallres}. The green, blue, and red lines refer to the case that no AP, 1 AP, and 2 APs are altered in reality. In all the cases, the system does not know how many APs are altered, needless to mention which one is altered.\\
\indent
The system exhibits high localization accuracy. Take the research laboratory as example. Compared with the baseline performance when the system knows that no AP is altered, the mean error increases by only 4.6\%, i.e., 0.29m in the `none' case. It shows that the system hardly mislabels unaltered APs as altered and that the localization accuracy remains high even when a few false alarms occur. As for the case that one or two APs are altered, the mean error rises by 5.4 cm and 8.6 cm respectively compared to the `none' case. The effectiveness of detection and fingerprint reconstruction can be easily verified.

%% file: 8_conclusion.tex
\section{Conclusion}
\label{sec: conclusion}
CSI based fingerprinting localization has attracted lots of interests. However, it has not been widely implemented on off-the-shelf smartphones.
In this paper we present CRISLoc, a system exploiting CSI as fingerprints and automatically reconstructing CSI fingerprints for smartphone localization. We successfully extract CSI from Nexus 5 with 20 MHz bandwidth in 2.4 GHz without building successful connections. In the research laboratory, we use CSI efficiently with our novel matching rule EEKNN, which reduces the error by 21.3 percent compared to traditional WKNN. In addition, our system is able to detect the AP alternation by a novel algorithm, cluster-outlier joint approach, with F1-scores 0.893 and 0.827 when one and two APs are altered. Reconstructed by a transfer learning method, the mean error only rises by 5.4cm and 8.6cm. {\color{blue}Future works for CRISLoc may focus more on realistic influence such as device direction and moving issues. More exhaustive fingerprinting database regarding to each direction should be collected for its direction performance. Data from embedded sensors on smartphones can fuse with CSI for indoor localization.}

%% file: appendix.tex
\section*{Appendix: Additional Notations}

\begin{table}[htbp]
    \renewcommand*{\arraystretch}{1.4}
    \centering
    \begin{tabular}{|p{1cm}|p{7cm}|}
    \hline
    Notation & Definition           \\ \hline
    $\hat{p}$ \& $p_i$&Estimated location \& position of the $i$th neighbor\\ \hline
    $w_i$&The weight for the $i$th neighbor in WKNN and EEKNN\\ \hline
    $\mathbf{P} \& P_s$&The whole set of APs \& a subset of APs\\ \hline
    $\rho$&Radius of neighborhood in DBSCAN\\ \hline
    $MinPts$&The minimum number of neighbors as a core point in DBSCAN\\ \hline
    $r_0$&The threshold for the joint clustering-outlier approach \\ \hline
    $P_1 \& P_2$&The predicted class of altered \& unaltered APs\\ \hline
    $f_i$&The frequency of the $i$th AP as an altered AP\\ \hline
    $\eta$&Adaptive weight factor of variance in Jenks method\\ \hline
    $l_0$&The threshold of reliability level in sequential analysis\\ \hline
    $min\_seq$&The minimum number of samples in sequential analysis\\ \hline
    $max\_seq$&The maximum number of samples in sequential analysis\\ \hline
    $\kappa_i$&The neighbor portion of training point $i$ or the $i$th neighbor\\ \hline
    $k'$&The total neighbor portion for a testing point, by default one\\ \hline
    \end{tabular}
\end{table}

\section*{Appendix: DBSCAN algorithm}

\begin{algorithm}[!t!h]
\caption{DBSCAN Algorithm for AP Clustering}
\label{alg: dbscan}
{\bf Input:} a set of objects\\
{\bf Parameter:} $\rho$: radius, $MinPts$: the minimum number of neighbors within $\rho$ (including itself).\\
{\bf Output:} cluster labels
\begin{algorithmic}[1]
\State label all objects as unvisited
\While{unvisited objects exist}
\State randomly select an unvisited objects $p$
\If{$p$ has $MinPts$ neighbors within $\rho$}
\State create a new cluster $C_i$ and add $p$ into $C_i$
\State create a new queue $Q_i$
\State add all neighbors of $p$ within $\rho$ into $Q_i$
\For {$q$ in $Q_i$}
\If {$q$ is unvisited \textbf{or} labelled as an outlier}
\State add $q$ into $C_i$
\If {$q$ has $MinPts$ neighbors within $\rho$}
\State add these neighbors into $Q_i$
\EndIf
\EndIf
\EndFor
\Else
\State label $p$ as an outlier
\EndIf
\EndWhile
\end{algorithmic}
\end{algorithm}

DBSCAN identifies clusters by the following law: given two global parameters radius $\rho$ and the threshold number of neighbors $MinPts$, a point is a member of a cluster either it is a core point that has at least $MinPts$ neighbors with the distances less than $\rho$, or it is a neighbor of a core point; otherwise it is considered as an outlier. The process to extend clusters in DBSCAN is based on breadth first search (BFS). 
Each unvisited point $p$ is tested to be either a member of a cluster or an outlier henceforth. If $p$ belongs one cluster, a new cluster $C_i$ is created and the above procedure is repeated for $p$'s neighbors until all the points have been visited. 
The DBSCAN algorithm is tailored for our 
localization system shown in Algorithm \ref{alg: dbscan}.\\

The parameter $MinPts$ is kept at the default value, which is the double of the dimension of the data space, i.e., four in two-dimension space. As for $\rho$, its optimal value varies in different situations, e.g., different number of altered APs. Instead of setting $\rho$ as a constant, we automatically select its value according to $k$-distance \cite{DBSCAN-k-distance}. $k$-distance means the distance of the $k$th nearest samples
. In our implementation, $k$ equals three. The sorted $k$-distance of localization results increase dramatically after a certain point. The results whose $k$-distance are smaller than that point probably lie in clusters. Therefore, we find the point where the slope starts to be steep and let its $k$-distance be the parameter $\rho$. With this approach, we avoid the risk of choosing the classification parameters arbitrarily. \\